\documentclass[9pt]{article}

\usepackage[T1]{fontenc}
\usepackage[english]{babel}
\usepackage[latin1]{inputenc}
\usepackage[charter]{mathdesign}
\usepackage[scaled=0.9]{helvet}
\usepackage{url,color}
\usepackage{enumerate}
\usepackage{verbatim}  

\usepackage[final]{graphicx}
\usepackage{subfigure} 
\usepackage{float}     

\usepackage{caption}
\usepackage[numbers,sort&compress]{natbib}

\usepackage{icomma}                        
\usepackage[bookmarks=false,pdfpagelabels,plainpages=false]{hyperref}  

\definecolor{uni}{rgb}{0.490,0.604,0.667}
\definecolor{ippblue}{cmyk}{1,0.45,0.04,0.}

\hypersetup{pdftitle={Multi-mode Alfv{\'e}nic Fast Particle Transport and Losses: Numerical vs.\ Experimental Observation},
            pdfauthor={Mirjam Schneller},
            pdfsubject={Nonlinear alfv{\'e}nic fast particle transport and losses},
            pdfcreator={\LaTeX2e},
colorlinks, linkcolor=black, urlcolor=blue, citecolor=green}

\begin{document}


\newcommand{\sref}{sec.\ \ref}    
\newcommand{\fref}{fig.\ \ref}    
\newcommand{\Fref}{Fig.\ \ref}    
\newcommand{\eref}{Fig.\ \ref}    

\newcommand{\qp}{$q$ profile}
\newcommand{\bef}{$\beta_{\mathrm{fp}}$}
\newcommand{\pitch}{\lambda}
\newcommand{\gyrad}{\rho}

\textbf{\LARGE{Multi-mode Alfv{\'e}nic Fast Particle Transport and Losses:\newline Numerical vs.\ Experimental Observation}}\\\\

\textbf{M. Schneller$^1$, Ph. Lauber$^1$, R. Bilato$^1$, M. Garc{\'i}a-Mu{\~n}oz$^1$, M. Br{\"u}dgam$^1$, S. G{\"u}nter$^1$ and the ASDEX Upgrade Team}\\
$^1$ Max-Planck-Institut f{\"u}r Plasmaphysik, EURATOM Association, Boltzmannstr. 2, D-85748 Garching, Germany\\
E-mail: \texttt{mirjam.schneller@ipp.mpg.de}

\begin{abstract}
In many discharges at \textsc{Asdex} Upgrade fast particle losses can be observed due to Alfv{\'e}nic gap modes, Reversed Shear Alfv{\'e}n Eigenmodes or core-localized Beta Alfv{\'e}n Eigenmodes. For the first time, simulations of experimental conditions in the \textsc{Asdex} Upgrade fusion device are performed for different plasma equilibria (particularly for different, also non-monotonic $q$ profiles). The numerical tool is the extended version of the \textsc{Hagis} code \cite{Pinches98,mwb_phd}, which also computes the particle motion in the vacuum region between vessel wall in addition to the internal plasma volume. For this work, a consistent fast particle distribution function was implemented to represent the strongly anisotropic fast particle population as generated by \textsc{Icrh} minority heating. Furthermore, \textsc{Hagis} was extended to use more realistic eigenfunctions, calculated by the gyrokinetic eigenvalue solver \textsc{Ligka} \cite{Lauber07}. The main aim of these simulations is to allow fast ion loss measurements to be interpreted with a theoretical basis. Fast particle losses are modeled and directly compared with experimental measurements \cite{Garcia10}. The phase space distribution and the mode-correlation signature of the fast particle losses allows them to be characterized as prompt, resonant or diffusive (non-resonant). The experimental findings are reproduced numerically. It is found that a large number of diffuse losses occur in the lower energy range (at around 1/3 of the birth energy) particularly in multiple mode scenarios (with different mode frequencies), due to a phase space overlap of resonances leading to a so-called domino \cite{Berk95-III} transport process. In inverted $q$ profile equilibria, the combination of radially extended global modes and large particle orbits leads to losses with energies down to 1/10th of the birth energy.
\end{abstract}

\section{Introduction}
    Fast particles are present in magnetic fusion devices due to external plasma heating and eventually due to fusion born $\alpha$-particles. It is necessary that these super-thermal particles are well confined while they transfer their energy to the background plasma. Fast particle populations can interact with global electromagnetic waves, leading to the growth of MHD-like and kinetic instabilities -- e.g.\ Toroidicity-induced Eigenmodes (TAE) \cite{Cheng85, Cheng86}, Reversed Shear Alfv{\'e}n Eigenmodes (RSAE) \cite{Berk01,Breizman03} or Beta-induced Alfv{\'e}n Eigenmodes (BAE) \cite{Heidbrink93, Turnbull93}. This in turn redistributes the fast ions, enhances fast ion losses and affects the confinement. As a consequence, fusion power and fuel is reduced and the machine wall may suffer damage.\\

    \textsc{Asdex} Upgrade (\textsc{Aug})\cite{Gruber84,GarciaRev09} and other machines of similar size (e.g.\ \textsc{DIII-D} \cite{zeeland11}), have the advantage to directly measure the fast ion losses due to core-localized MHD activity, since the particle orbits are relatively large compared to the machine size. The fast ion loss detector (\textsc{Fild}) at \textsc{Asdex} Upgrade provides energy and pitch angle resolved measurements of fast ion losses \cite{GarciaRev09,Garcia10}.\\

To find out about the loss mechanisms, loss features such as phase space pattern and ejection frequency are analyzed. Both can help to learn about whether or not the losses are caused by mode-particle (double) resonance (\emph{resonant losses}), phase space stochastization (\emph{diffusive losses}) \cite{Sigmar92} or none of both. Therefore, the losses are categorized: all particles that are unconfined solely due to their large orbit width -- caused by a high birth energy -- or a birth position radially far outside, are called \emph{prompt losses}. Their ejection frequency is not related to any of the mode frequencies, and the losses are therefore inherently \emph{incoherent}. Losses that are caused due to wave-particle interaction can be both, either \emph{coherent losses}, or incoherent as well. They are coherent, if they show a correlation with one of the mode frequencies or the beat frequency of different modes, but incoherent if they are ejected e.g.\ due to phase space stochastization. To learn about the role of the modes in the ejection mechanisms, the losses are further studied with respect to the number and type of modes present in the plasma and the \qp\ (as this determines particle orbit widths and also mode growth). Eventually, this work aims to reproduce quantitatively the pitch and energy distribution and qualitatively the decomposition into \emph{coherent} and \emph{incoherent} losses of the  of the measured fast ion losses \cite{Garcia10} at the different times in the \textsc{aug} discharge \#23824. From the code point of view, the study can be regarded as a code validation against the experiment, which is possible in small machines with large particle orbits (due to \textsc{Icrh}) such as \textsc{Asdex} Upgrade, where a large number of losses appears at the first wall and the \textsc{fild}.\\

This paper is organized as follows: \sref{sec:experiment} presents the experimental settings the numerical simulations are based on, as well as experimental loss observation. In \sref{sec:hagis}, the \textsc{Hagis} code is shortly introduced, especially the importance of the vacuum extension \cite{mwb_phd} is demonstrated. Next (\sref{sec:resplot}), the resonant and loss areas in phase space are given for the scenario that is studied. In \sref{sec:numloss}, the advance towards more realistic simulations is documented, followed by the presentation of the final results.

\section{Experimental Observation}\label{sec:experiment}
    The \textsc{Icrh} minority-heated \textsc{aug} discharge \#23824 is especially suitable for the investigation of fast particle-wave interaction and losses, as it is characterized by an inverted \qp\ after the current ramp-up phase, which relaxes to a monotonic profile as $q$ decreases (see \fref{q23824}). As particle confinement depends largely on the \qp\ due to the proportionality between orbit width and $q$ value, the inverted \qp\ with its higher absolute $q$ values in the core is expected to impact the amount of losses and also the wave-particle interaction.\\
    The simulations are based on two different MHD equilibria -- at $t=1.16$\ s and $t=1.51$\ s. The total plasma current is constant between both time points ($I=800$\ kA), only the current density penetrates inwards. Consequentially the inverted \qp\ at the earlier time point changes to a monotonic \qp\ at the later time point.

\section{The Hagis Code}\label{sec:hagis}
    The numerical investigations presented in this work are performed with the \textsc{Hagis} code \cite{Pinches98, sip_phd}, a nonlinear, driftkinetic, perturbative Particle-in-Cell code (current release 12.05). \textsc{Hagis} models the interaction between a distribution of energetic (fast) particles and a set of Alfv{\'e}n eigenmodes. It calculates the linear growth rates as well as the nonlinear behavior of the mode amplitudes and the fast ion distribution function that are determined by kinetic wave-particle nonlinearities. It is fully updated to work with MHD equilibria given by the recent \textsc{Helena} \cite{Huysmans91} version.\\
    The plasma equilibrium and in particular the \qp\ is determined by Alfv{\'e}n spectroscopy of the RSAEs: magnetic pick-up coil data and soft X-ray emission measurements \cite{Igochine03} are used to determine the \qp\ minimum value and location. The plasma equilibrium for \textsc{Hagis} is based on the \textsc{Cliste} code \cite{Carthy12}, then transformed via \textsc{Helena} to straight field line coordinates and to Boozer coordinates via \textsc{Hagis}.

    \begin{minipage}{1\textwidth}
      \begin{minipage}{0.6\textwidth}
        \begin{figure}[H]
          \centering
          \includegraphics[width=0.8\textwidth, height=5.5cm]{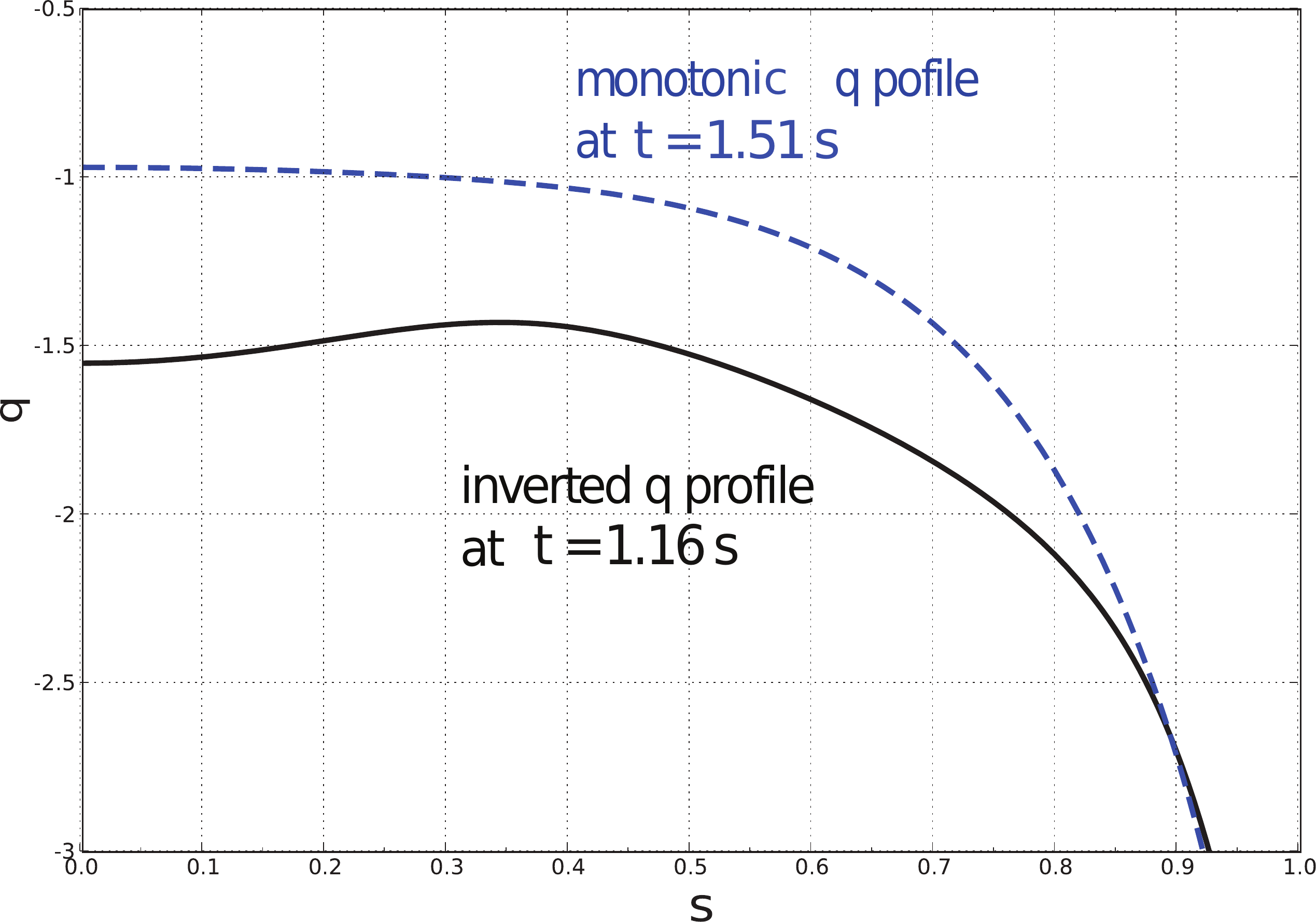}
          \caption{\itshape \qp\ of \textsc{aug} discharge \# 23824 at different times,
            obtained from \textsc{Cliste} calculation constraint by
          Alfv{\'e}n spectroscopy measurements as described in detail in ref.\ \cite{Lauber09}.
          The black solid line shows the \qp\ at $t=1.16$\ s, whereas the blue dashed line refers to 
          $t=1.51$\ s:
          the \qp s differ in the shape (the earlier one is inverted, the later one monotonic), but
          also in the absolute values (the inverted \qp\ has higher absolute $q$ values). Note: $q < 0$ due 
          to \textsc{aug}'s helicity of the magnetic field.
          }
          \label{q23824}
        \end{figure}
      \end{minipage}
      \hfill
      \begin{minipage}{0.35\textwidth}
        \begin{figure}[H]
           \hspace{1cm}\includegraphics[width=0.9\textwidth, height=7cm]{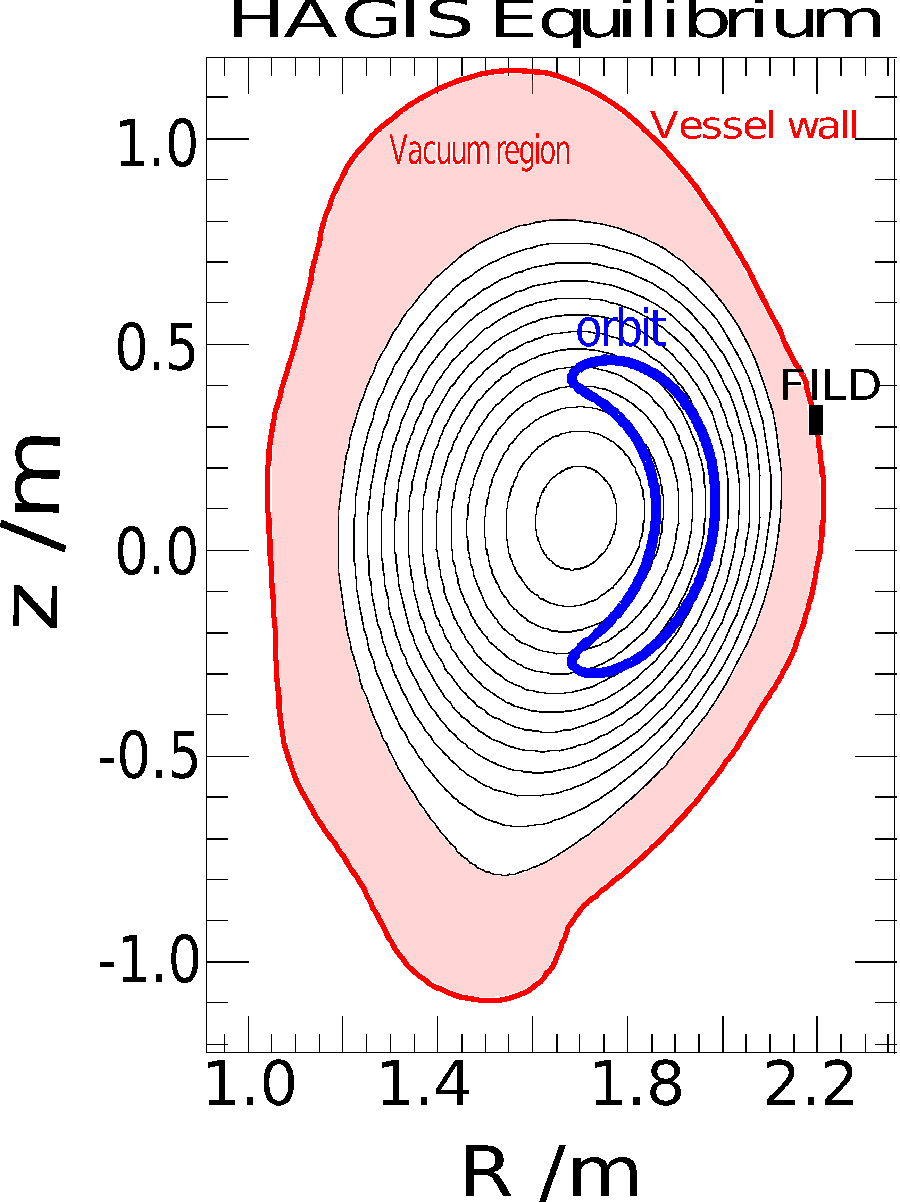}
           \caption{\itshape Hagis equilibrium (flux surfaces in black) with vacuum region (red), 
             representative banana orbit (blue) and \textsc{Fild} position (black square).}
           \label{hagis_equilibrium}
        \end{figure}
        \vspace{0.5cm}
      \end{minipage}
    \end{minipage}
    \newpage
    Recently, \textsc{Hagis} has been extended to the vacuum region (see \fref{hagis_equilibrium}), making it possible to track particles that re-enter the plasma \cite{mwb_phd}, and to investigate losses including finite Larmor radius effects. To allow for the second point within the frame of a drift kinetic ansatz, a tube around a particle's guiding center is assumed, whenever it approaches the wall. Where this tube intersects the wall, the particle is lost. For scenarios with many lower energetic losses, as they appear e.g.\ in all inverted \qp\ simulations presented in this work, the use of the vacuum extension is crucial. \Fref{run0502+508_Eloss} visualizes the difference in the losses' energy spectra obtained with (black curve) and without (red curve) the vacuum extension -- if simulating without it, the lower energetic losses are massively overestimated. The overestimation results from to the fact, that particles are considered lost, as soon as they cross the separatrix. However, particles with very small gyroradii (i.e.\ low perpendicular energies), often re-enter the plasma and are not lost. The overestimation of lost markers if simulating without the vacuum extension is over 130\% in the given example. \textsc{Icrh}-generated fast particles are characterized by relatively low pitches $\pitch = v_\|/v$, thus, they have low perpendicular energies only if their total energy is in a lower range. These particles have small gyroradii, and therefore a higher probability to re-enter the plasma. Further, the distance between the separatrix and the first wall is important, which depends on the given equilibrium in the particular fusion device.
    The vacuum extension allows particles to travel through the vacuum and checks if the particle hits the machine specific first wall through assuming a `tube' around its guiding center with the radius of the gyroradius.  Simulations including the vacuum region are therefore computationally more expensive (roughly between 120\% and 300\% compared to the simulations without the extended version).

    \begin{figure}[H] 
      \centering
      \includegraphics[width=0.7\textwidth, height=3cm]{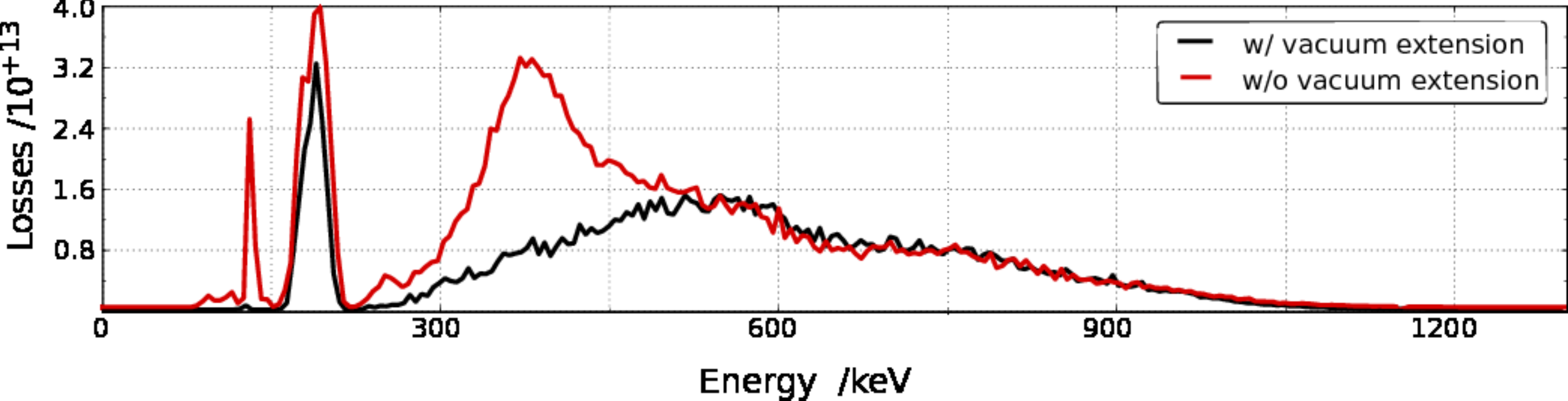} 
      \caption{\itshape Energy spectra of losses simulated in the inverted \qp\ 
        with the eigenmodes given by \textsc{Ligka}\ using \textsc{Hagis} with 
        its vacuum extension 
        (black curve) and without (red). The amplitudes for these simulations were fixed
        at $\delta B/B = 5.1\cdot 10^{-3}$. The losses shown appeared in a time interval starting 
        after approximately 10 RSAE wave periods ($t \in [0.2,1.5] \cdot 10^{-3}$\ s) to 
        avoid the prompt losses.} 
      \label{run0502+508_Eloss}
    \end{figure}
    \begin{figure}[H]
        \centering   
        \subfigure[\itshape at $t = 1.16$\ s -- inverted \qp]{\includegraphics[width=0.45\textwidth,height=5.5cm]{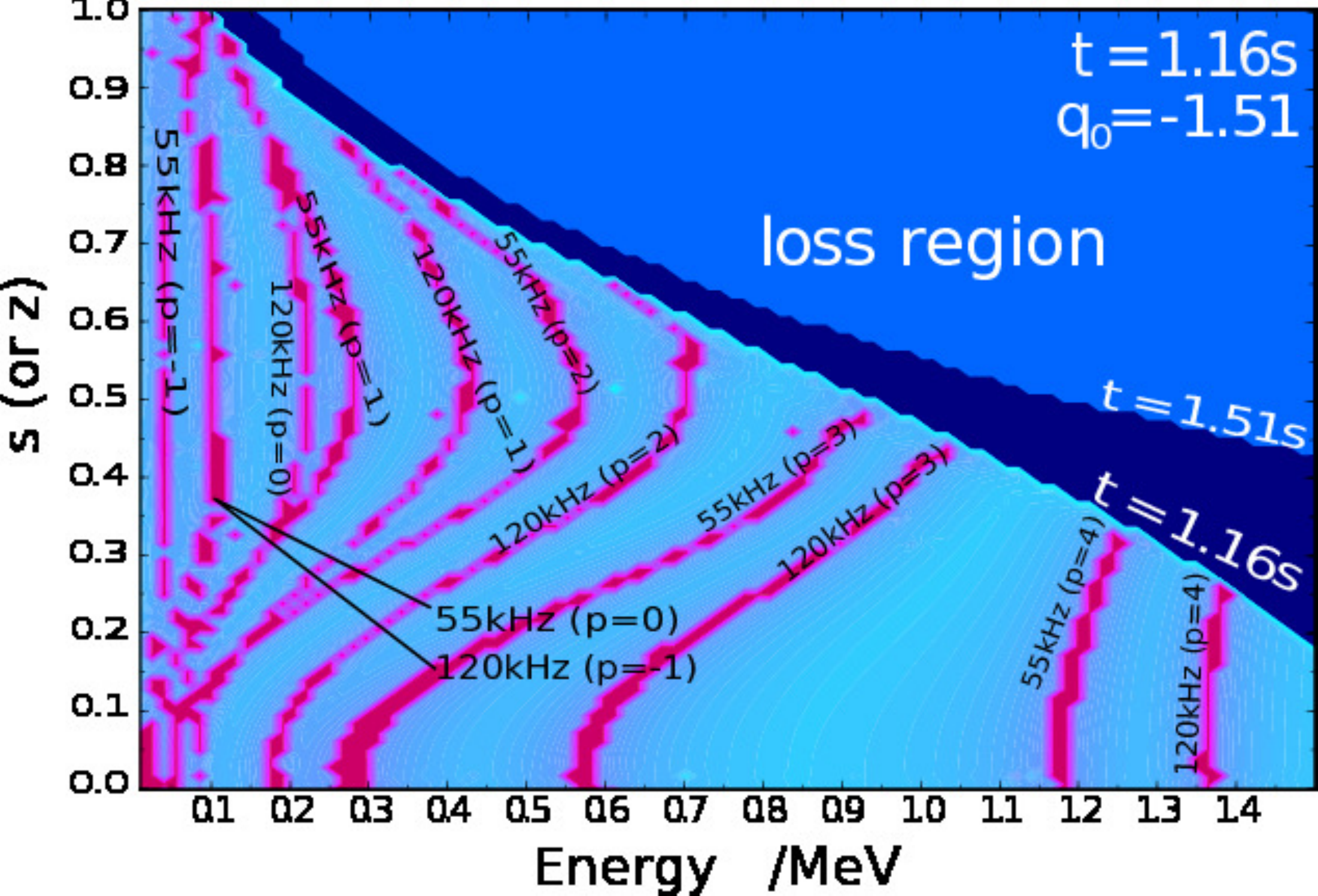}}\hspace{1cm} 
        \subfigure[\itshape at $t = 1.51$\ s -- monotonic \qp]{\includegraphics[width=0.45\textwidth,height=5.5cm]{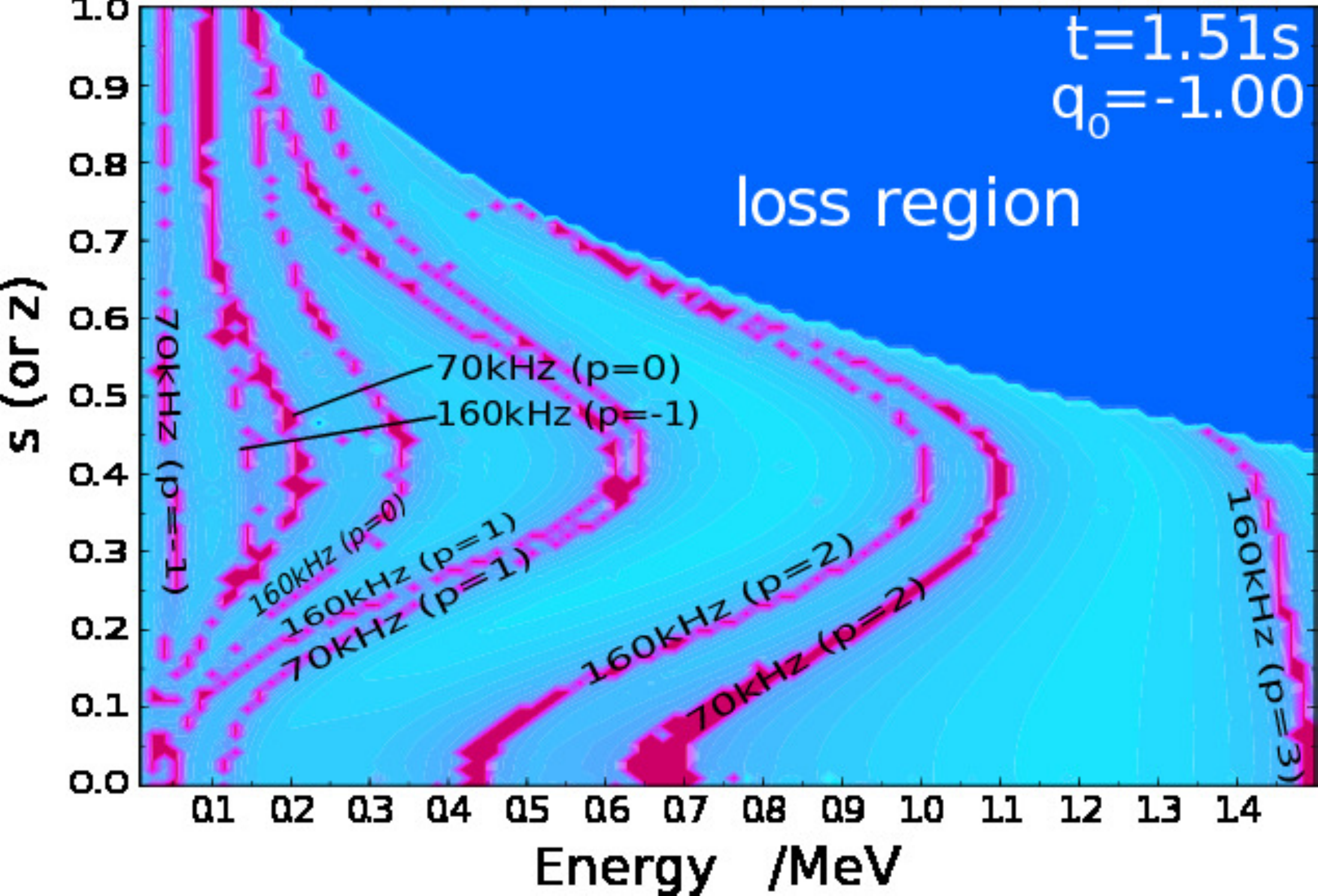}} 
        \caption{\itshape Resonance plots produced by the extended version of \textsc{Hagis} for the equilibria of \textsc{aug} discharge \#23824 at two different time points with different \qp s. The vertical axis gives the radial position of the particle's bounce point (where pitch $\pitch =0$), the horizontal axis the energy. Pink indicates resonant areas, with respect to the main MHD modes as diagnosed experimentally ($n=5$ TAE of $120$\ kHz, $n=4$ RSAE of $55$\ kHz at $t = 1.16$\ s and $n=5$ TAE of $160$\ kHz, $n=4$ BAE of $70$\ kHz at $t = 1.51$\ s). Particles in the blue regions are lost. The loss region of the right picture is plotted superimposed in the left image to point out the difference.}
        \label{resplot_AUG-23824}
    \end{figure}

\section{Resonances and Loss Regions}\label{sec:resplot}
    The \textsc{Icrh} minority-heating produces mainly trapped particles with their bounce points within the heating region that is at constant major radius above and below the magnetic axis (see \fref{loading-weighting}b). Trapped particles with a bounce frequency $\omega_{\mathrm{b}}$ and toroidal precession frequency $\omega_{\mathrm{tp}}$ interact with MHD modes of a certain frequency $\omega$, if the \emph{resonance condition}, $\omega - n\omega_{\mathrm{tp}}-p\omega_{\mathrm{b}} \approx 0$ is fulfilled, where $n$ is the toroidal mode harmonics and $p$ the particles' bounce harmonics \cite{porcelli94}. Plotting the left side of the resonance condition as a contour plot over the particle energy $E$ and the $z$ coordinate of the particle's bounce point over the magnetic axis gives the \emph{resonance plot} \cite{Pinches06} shown in \fref{resplot_AUG-23824}. Since at the bounce point, the pitch is $\pitch \equiv v_\|/v = 0$, the second dimension in energy space is kept fixed. The normalized $z$ coordinate can be translated to the radial coordinate $s$, the square root of the normalized poloidal flux: $s=(\psi_{\mathrm{pol}}/\psi_{\mathrm{pol, edge}})^{1/2}$. The darker blue areas give the \emph{loss regions}: in the respective phase space particles are unconfined.\\
   The loss region is much broader in the earlier case, with the inverted \qp. The reason for the decrease of the loss region is the decreasing $q$ in the plasma center region, leading to a decrease of orbit width (due to width $\propto q$) and therefore better particle confinement.

\section{Fast Particle Losses in Numerical Simulations vs.\ Experimental Measurements}\label{sec:numloss}

  \subsection{Experimental Loss Measurements}
    For the \textsc{aug} discharge \#23824, the experiment gives a loss signal over time as depicted in figure 4 of ref.\ \cite{Garcia10}. The upper part shows the spectrogram of that data, which allows to identify the mode frequencies of the Alfv{\'e}n Eigenmodes. One can distinguish a large signal with a vast majority of incoherent losses at earlier time points, whereas after $t = 1.4$\ s the signal is very low, consisting of coherent losses only. A Fourier analysis revealed two types of losses: \emph{coherent} losses, i.e.\ losses ejected at a frequency correlated with the MHD mode frequency on and \emph{incoherent} losses, that do not show such a correlation.

  \subsection{Realistic Simulation Conditions}\label{sec:simcond}
      The fast particle population is simulated as protons (concerning mass and charge), according to how it is created by \textsc{Icrh} minority heating in a deuterium plasma \cite{Garcia10}. The volume averaged fast particle beta is set to \bef\ $\approx 0.05$\%. This can be estimated from the construction of the MHD equilibrium and different kinetic plasma profile measurements. However, the \textit{volume averaged} \bef\ does not alone determine the radial gradient of the distribution function, which governs the mode drive. For the most realistic modeling in the following sections, the \bef\ value given above is reduced to \bef\ $= 0.02$\% to obtain mode amplitude saturation levels comparable to the experimental values. However, the larger \bef\ value leads to a faster mode growing and slightly overestimated amplitude saturation levels. Since we are not interested in the total amount of losses but rather in their energy and pitch distribution as well as their nonlinear behavior (which is the same for both \bef), in the indicated simulations, the larger \bef\ is chosen to save computational costs. With either beta value, the nonlinear regime is steady-state without frequency chirping, since there is no damping involved.\\

\begin{figure}[H]
        \centering
         \subfigure[\itshape radial direction $s$]{\includegraphics[width=0.4\textwidth,height=3.5cm]{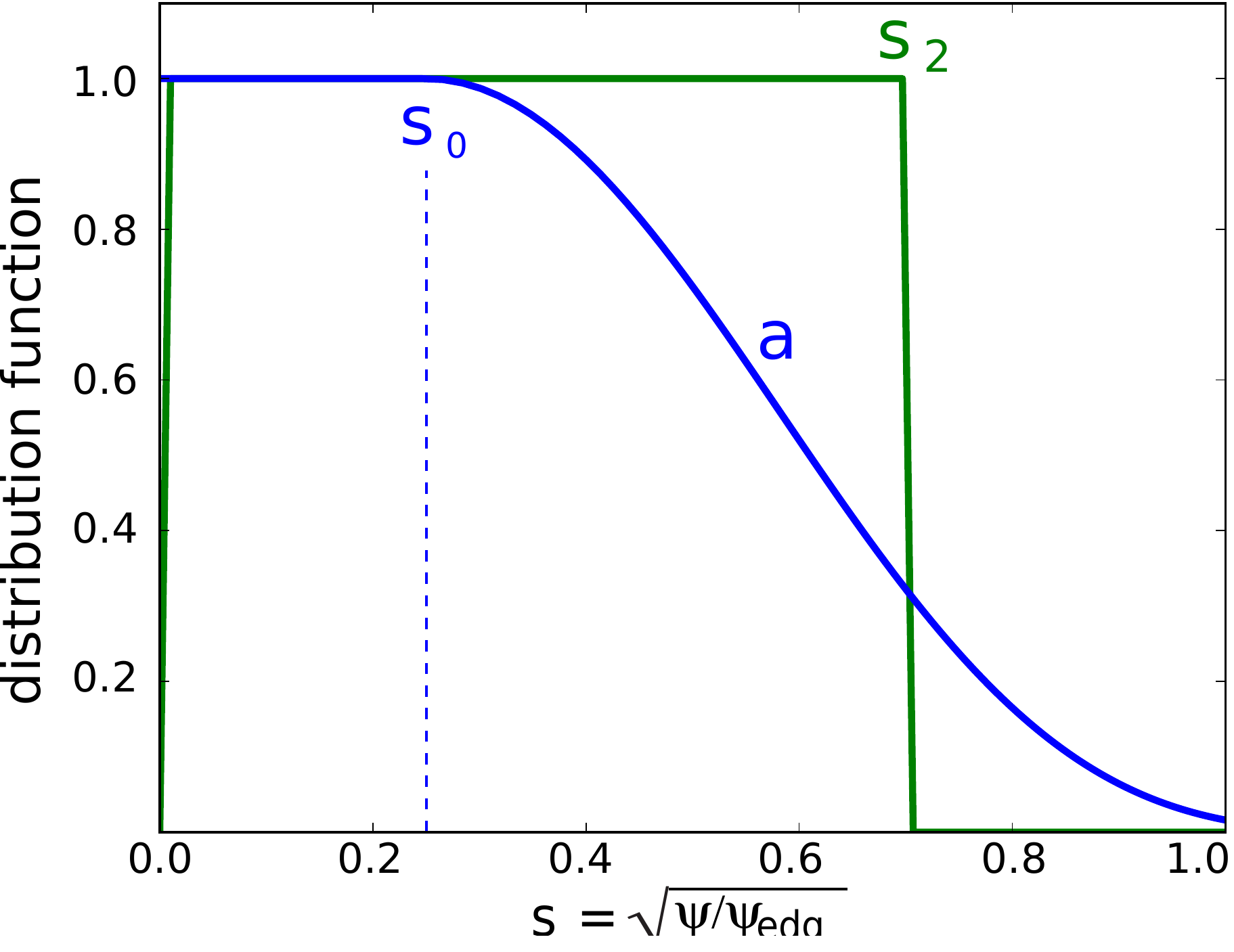}}
         \hspace{2cm}\subfigure[\itshape loading along $s$ and $\theta$]{\includegraphics[width=0.3\textwidth,height=4.5cm]{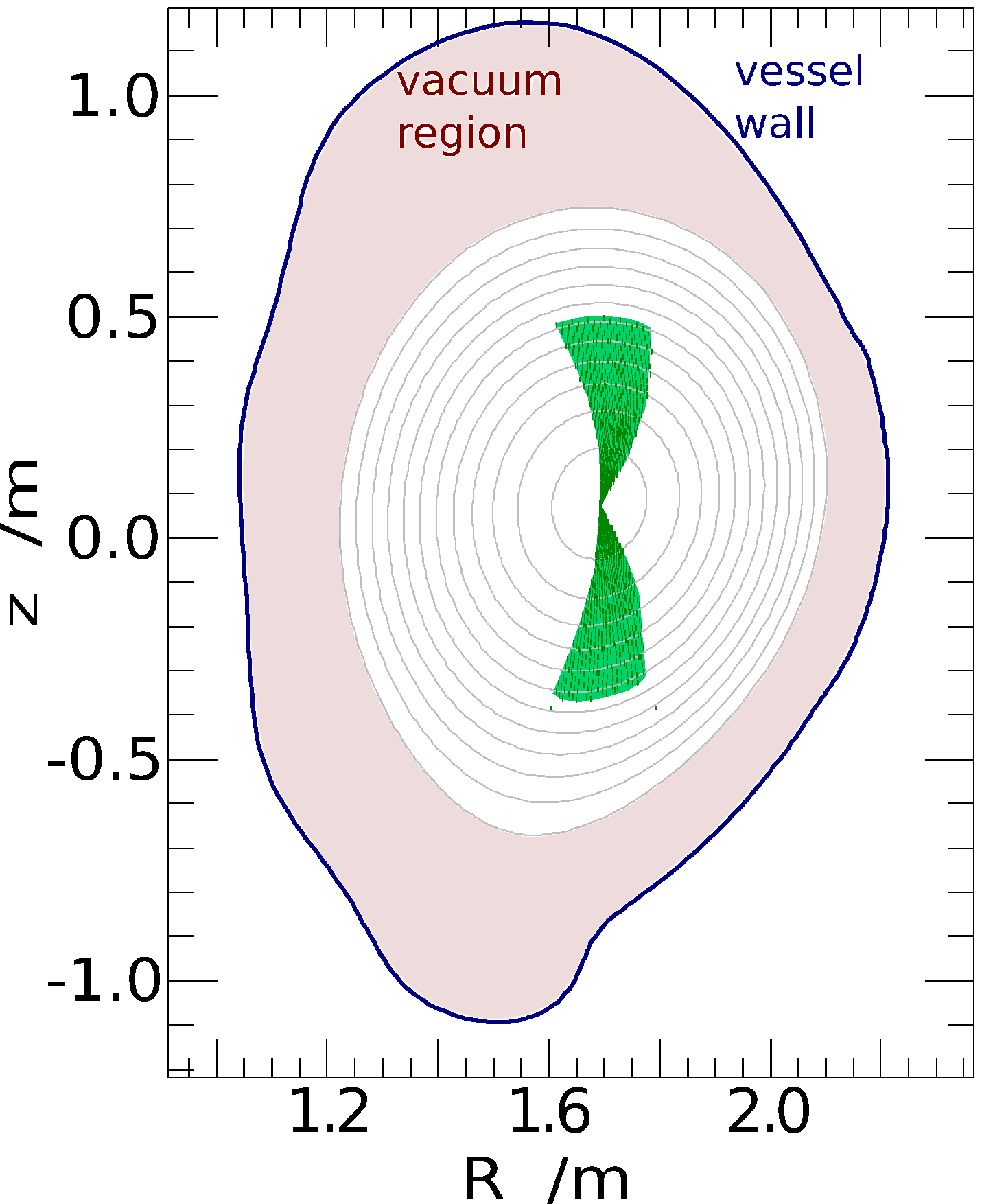}}\hspace{0.75cm} 
         \caption{\itshape Regions where markers are loaded (green) in the different phase space 
           directions and the distribution function for the unperturbed part $f_0$ (``weighting'', blue, only in (a)).}
        \label{loading-weighting}
      \end{figure}

    To be able to simulate the interaction of these \textsc{Icrh}-generated fast ions with Alfv{\'e}nic waves, the distribution function in \textsc{hagis} was adapted (see also ref.\ \cite{mirjam_phd}): especially the anisotropy in the pitch $\pitch$ and orbit topology is taken into account: \textsc{Icrh}-generated fast ions are characterized by low pitches and are mainly trapped (banana orbits). The major challenges are the following: first, and different from \textsc{Nbi}-heated plasmas, the fast ion distribution function in \textsc{Icrh}-heated plasmas is almost unknown experimentally (especially for this discharge). Second, the strong anisotropies of the distribution function are in coordinates that are not constants of the motion (pitch $\pitch$ and poloidal angle $\vartheta$ -- coordinates of velocity and real space that change along the particle orbit), making it difficult to evolve such a distribution function in \textsc{Hagis}. Therefore, in the new implementation, the weighting is not performed in $\pitch$ and $\theta$ anymore, but in coordinates that are constants of the motion. The quantity $\Lambda$ fulfills this condition: it is connected to the pitch $\pitch$, but in contrast to that, is not subject to changes along the orbit:
    \begin{equation}\label{bigLambda}
      \Lambda ~:=~ \frac{\mu~B_\mathrm{mag}}{E_\mathrm{tot}} ~=~ \frac{B_\mathrm{mag}}{B} \left(1-\pitch^2\right),
    \end{equation}
    where $B_\mathrm{mag}$ is the magnetic field at the magnetic axis, introduced as normalization, $B$ is the magnetic field at the respective position $B(s,\theta)$ and $\pitch=\pitch(s,\theta)$ the pitch at this position. Due to the necessity of transforming $\pitch$ coordinate into the constant of the motion $\Lambda$, one obtains one velocity space coordinate dependent on the spacial coordinates: $\Lambda(s,\theta) = B_\mathrm{mag}/B(s,\theta)(1-\pitch^2)$.
    The distribution in $\Lambda$ is determined by marker loading in poloidal angle ($\theta$) and pitch ($\pitch$) space. Unless otherwise noted, markers start at pitches and poloidal angles of
    \begin{equation}\label{formula:lampollim}
       \pitch \in [0 \pm 0.2] \mathrm{~~~~~~~and~~~~~~~} \theta \in [\pm 90^{o} \pm 17.2^{o}].
    \end{equation}
    A poloidal cut of the two initial cones, where the \textsc{Icrh}-generated fast particles are loaded is shown in \fref{loading-weighting}b.
    A reasonable $\Lambda$ weighting function is the modified Gaussian
    \begin{equation}\label{formula:Lambdis}
      f(\Lambda) = \exp{\{((\Lambda - \Lambda_0)/\Delta \Lambda^{0.2})^2\}},
    \end{equation}
    where $\Lambda_0$ and $\Delta \Lambda$ result from $\lambda_0$ and $\Delta \lambda$ as well as from the (unperturbed) magnetic field  $B(s,\theta)$ (according to \eref{bigLambda}) and are thus position-dependent.
    The chosen $\Lambda$ weighting results in a relatively broad $\Lambda$-distribution, which is realistic when considering also pitch angle scattering, but too broad to represent the instantaneous effect of the \textsc{Icrh} heating source. Therefore, the pattern of the losses appearing at the very beginning (prompt losses) is broader than what would be realistic. To investigate this pattern, additional simulations are carried out with marker loading in 
    \begin{equation}\label{formula:lampollim2}
       \pitch \in [0 \pm 0.05] \mathrm{~~~~~~~and~~~~~~~} \theta \in [\pm 90^{o} \pm 5^{o}].
    \end{equation}

\begin{figure}[H]
        \centering
         \includegraphics[width=0.8\textwidth,height=3.5cm]{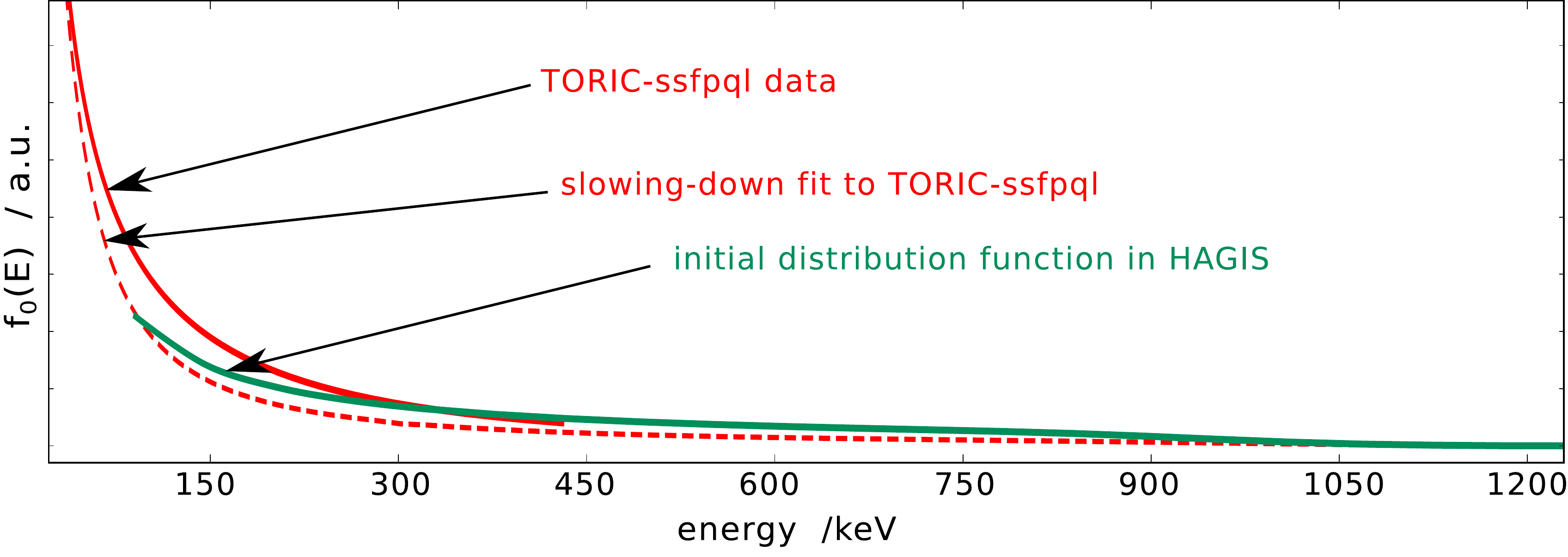}
         \caption{\itshape Energy distribution function in \textsc{Hagis} (green) and as given by the \textsc{Toric-ssfpql} code (red, which focuses on the lower energy range and reaches only up to 430\ keV). The red dashed lines give the slowing-down fit to the \textsc{Toric-ssfpql} data.}
        \label{energy-disb}
      \end{figure}

\begin{figure}[H]
      \centering
      \subfigure[\itshape Prompt losses ($t < 10^{-4}$)]{\includegraphics[width=0.43\textwidth,height=5.cm]{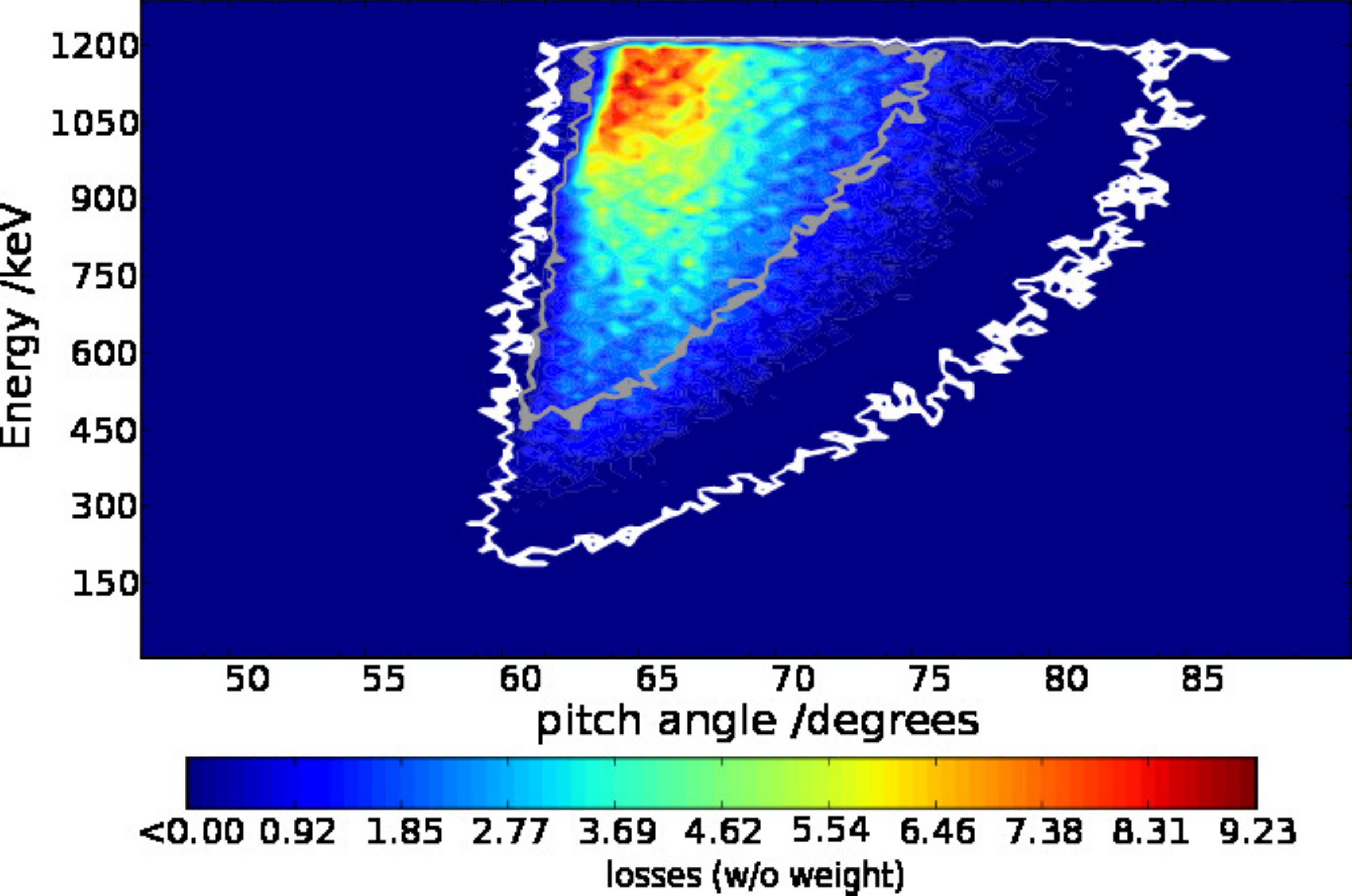}}
      \subfigure[\itshape Resonant/diffusive losses]{\includegraphics[width=0.43\textwidth,height=5.cm]{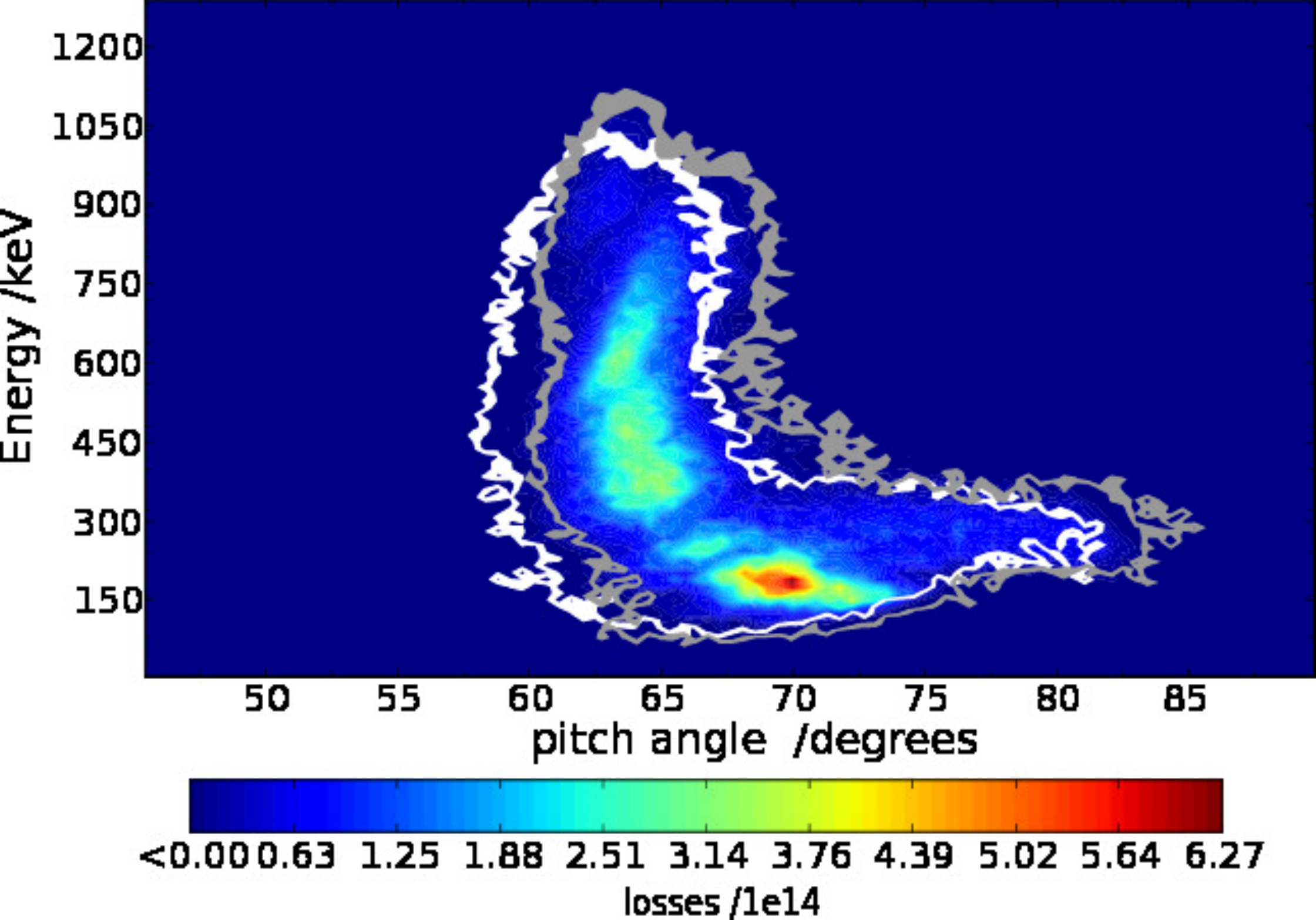}}
      \caption{\itshape Losses at first wall in pitch angle-energy ($\pitch^o$-$E$) space for three double-mode simulations in the inverted \qp\ (modes as explained below and shown in \fref{ligka-input}a, \bef=0.05\%). The simulations differ in the radial distribution function: the color code gives the prompt (a) and later (b) losses for the intermediate distribution function, the gray lines give the boundary of the pattern, when using the steeper radial distribution function; the white line gives the respective result for the flatter distribution function.}
      \label{run0447+452+456_pEloss}
    \end{figure}

\begin{figure}[H]
      \centering
      \subfigure[\itshape Prompt losses ($t < 10^{-4}$)]{\includegraphics[width=0.43\textwidth,height=5.cm]{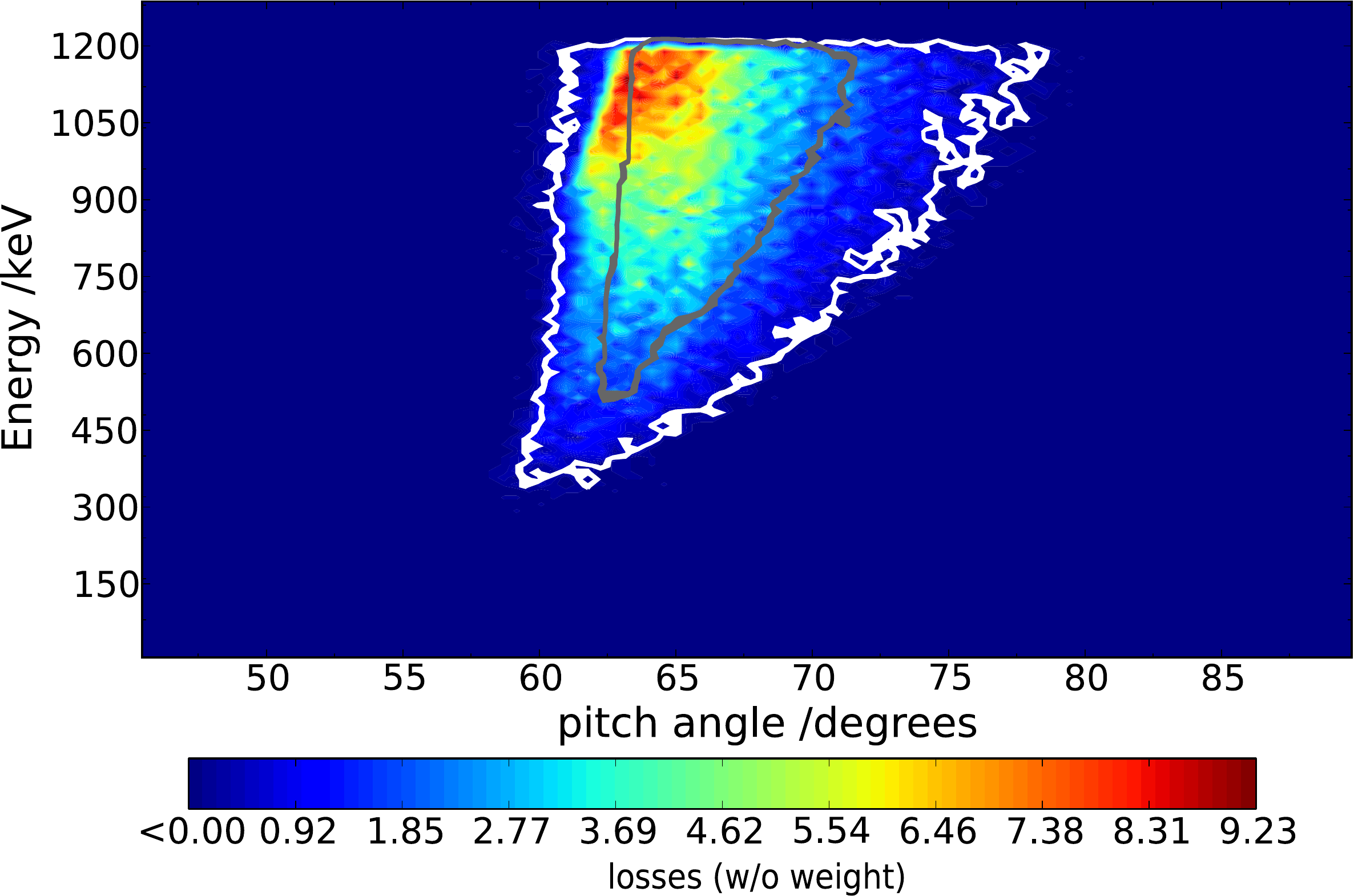}} 
      \subfigure[\itshape Resonant/diffusive losses]{\includegraphics[width=0.43\textwidth,height=5.cm]{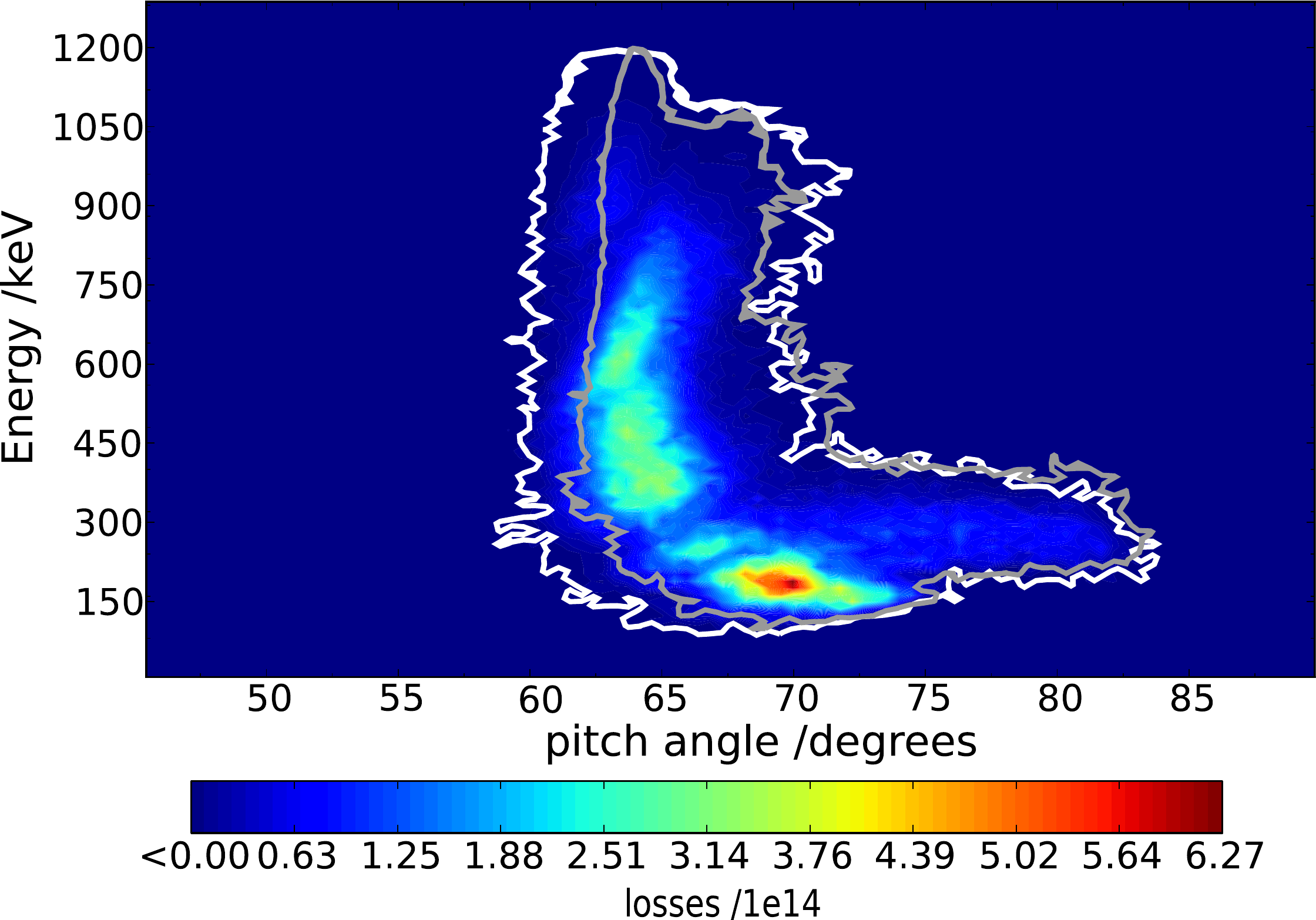}} 
      \caption{\itshape Losses at first wall in pitch angle-Energy ($\pitch^o$-$E$) space for two double-mode simulations in the inverted \qp\ (modes as explained later and shown in \fref{ligka-input}a, \bef=0.05\%). The simulations differ in the $\Lambda$ distribution function: the color code gives the prompt (a) and later (b) losses for the $\Delta \theta = \pm 17.2^o, \Delta \pitch = \pm 0.2$ marker loading, the white lines give the boundary of the pattern, when using $\Delta \theta = \pm 11.5^o, \Delta \pitch =\pm 0.2$ loading, the gray lines for $\Delta \theta = \pm 2.5^o, \Delta \pitch =\pm 0.05$.} 
      \label{run0447+471+472_pEloss}
    \end{figure}

      In radial position space, markers were loaded within $s_2 \leq 0.7$ (unless otherwise noted). The weighting function is implemented as constant for $s < s_0$ and drops down according to a Fermi-like potential law for $s \geq s_0$, with $s_0 = 0.3$ (unless otherwise noted), as shown in \fref{loading-weighting}a:
      \begin{equation}\label{formula:sdis}
        f(s) = \Bigg\{ \begin{array}{cc}
                          \mathrm{const.} ~~& \mathrm{for}~~ s < s_0\\
                          (1-(s-s_0)^2)^a   & \mathrm{for}~~ s \ge s_0,
                       \end{array}
      \end{equation}
     Concerning the energy distribution $f(E)$, a slowing-down function is used,
    \begin{equation}\label{formula:Edis}
      \nonumber f(E) = \frac{1}{E^{3/2}+E^{3/2}_{\mathrm{c}}}\mathrm{erfc}\left(\frac{E-E_0}{\Delta E}\right),
    \end{equation}
    with $E_\mathrm{c}=19.34$\ keV, $\Delta E=149.9$\ keV and $E_0=1.0$\ MeV, although a slowing-down function is not considered the most adequate distribution function for \textsc{Icrh} generated fast particles. However, since it is almost unknown experimentally for this discharge, the \textsc{Toric-ssfpql} code \cite{Bilato12} was consulted. The data in the highest energy range calculates by \textsc{Toric-ssfpql} (red line in \fref{energy-disb}), which corresponds to the medium and most relevant energy range in the presented investigations, can justify the use of a slowing-down function in the \textsc{Hagis} simulations (green line in \fref{energy-disb}). 
    Still, $f$ is independent in each dimension, $f= f_s(s) f_E(E) f_\Lambda(\Lambda)$. A comparison with the distribution function as calculated by the \textsc{Toric-ssfpql} code confirms this as a valid approximation for the treatment in $\Lambda$. Further, it is comparable to previous approaches, as reported in ref.\ \cite{Zonca00}. However, concerning radial-energy space, further improvements towards more realistic distribution functions, which are non-separable in the two dimensions ($f(s,E)$), are currently under investigation. An analytical model is given in ref.\ \cite{Troia12}, and will be implemented into \textsc{Hagis}, once its parameters are optimized numerically (by \textsc{Toric-ssfpql}) and validated by the comparison with experimental findings. A sensitivity investigation (presented in the following) confirms robustness of non-prompt losses against changes in the presently used distribution function:\\ 
A first scan is performed with three different radial distributions of decreasing steepness while the radial extent broadens -- the parameter $s_0$ in the weighting function \eref{formula:sdis} takes the values 0.15, 0.25, 0.35 (for all three, it is $a=5$) and the marker loading reaches out until $s_2 =0.6,~ 0.7$ and $0.8$ respectively. The second scan is carried out for three poloidal loadings according to $\Delta \theta = \pm 17.2^o, \Delta \pitch = \pm 0.2$ and $\Delta \theta = \pm 11.5^o, \Delta \pitch =\pm 0.2$ as well as $\Delta \theta = \pm 2.5^o, \Delta \pitch =\pm 0.05$. In \fref{run0447+452+456_pEloss}, the boundaries of the loss pattern are shown, as obtained for the scan in the radial distribution function. One can see, that the prompt losses (a) are relatively sensitive to the distribution function. However, they cannot be determined quantitatively within the used model anyhow. The reason lies inherently in the code model, which does not calculate the slowing-down process of the (unperturbed) fast particle distribution function. Since the simulated time is short compared to the slowing-down time, a stationary situation is considered, where heating and collisional dissipation balances each other. The non-prompt losses (b) in contrast, are quite robust against minor changes in the radial and poloidal distribution function. A similar result is obtained for the poloidal scan (see \fref{run0447+471+472_pEloss}).
   
\begin{figure}[H]
        \centering
        \subfigure[\itshape at $t = 1.16$~s: 55~kHz RSAE (gray) and 120~kHz TAE]{\includegraphics[width=0.49\textwidth, height=4cm]{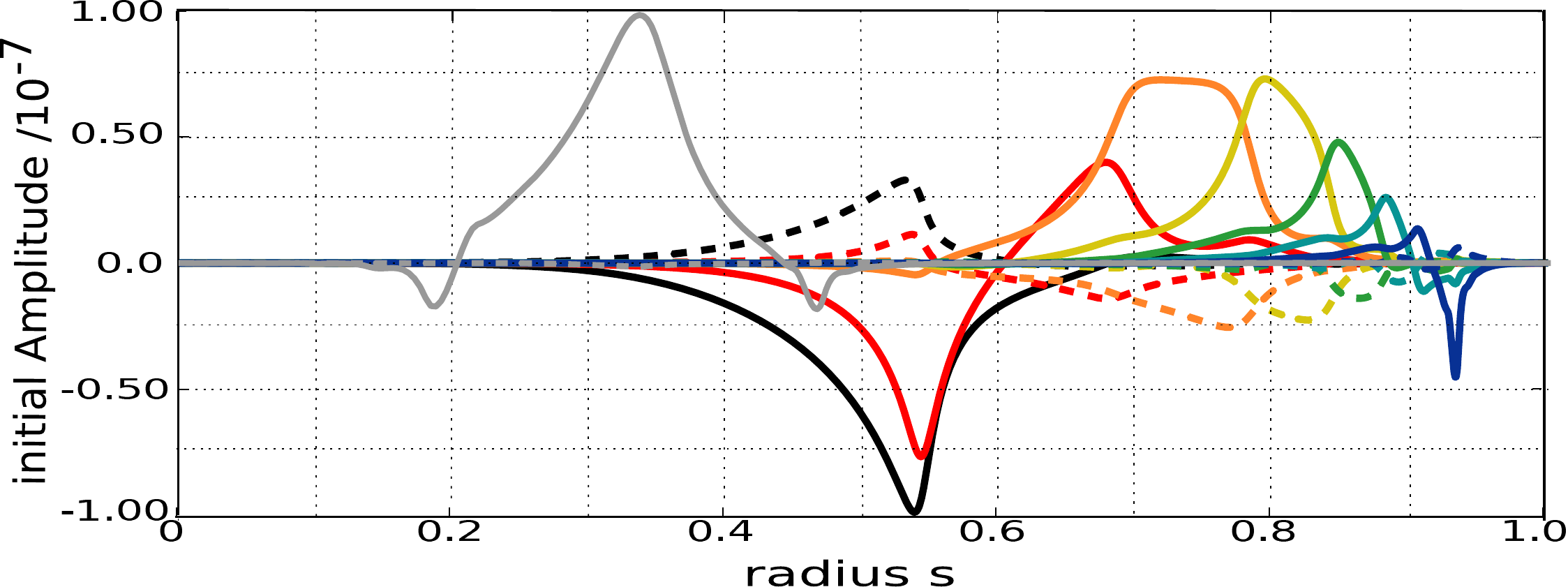}} 
        \subfigure[\itshape at $t = 1.51$~s: 70~kHz BAE (gray) and 160~kHz TAE]{\includegraphics[width=0.49\textwidth, height=4cm]{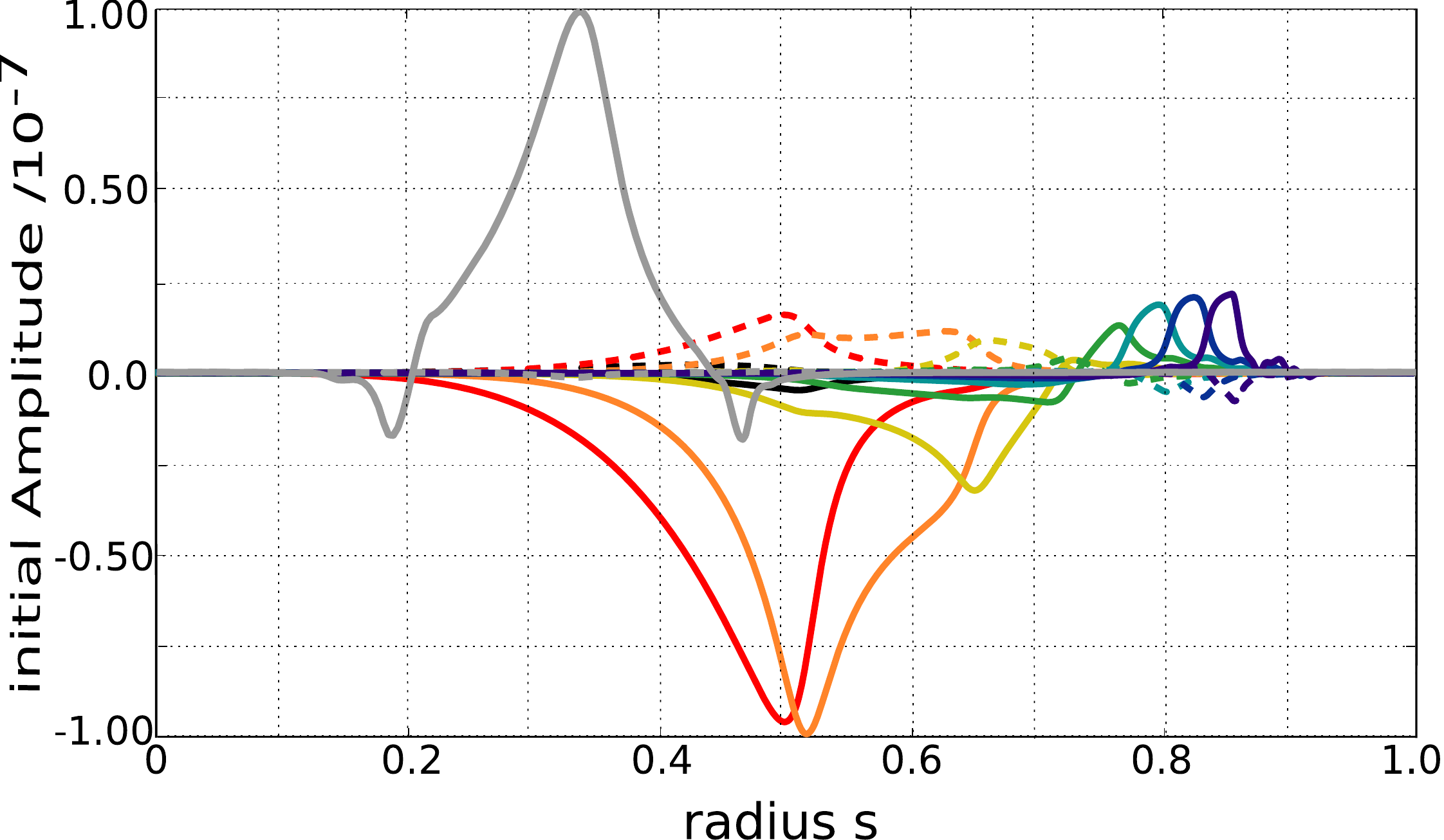}} 
        \caption{\itshape The radial structure including poloidal harmonics with $m=4...10$ (a) and $m=4..11$ (b) 
          (from black to blue/violet) of the $n=4$ TAE and the $m=4,~n=4$ RSAE of (a)/BAE (b) (gray) 
          as calculated by \textsc{Ligka} for the \textsc{aug} discharge \#23824 equilibrium at two 
          different time points.
          The given perturbation amplitude 
          refers to the real (solid lines) and the imaginary (dashed lines) part
          of the electric perturbation potential $\tilde{\Phi}$. The values are normalized, such that
          unity corresponds to the initial value of $\delta B/B = 10^{-7}$.}
        \label{ligka-input}
     \end{figure}

    Having adapted the simulation conditions towards more realistic fast particle distribution functions, the input perturbation data is improved (see also ref.\ \cite{mirjam_phd}). Based on the resulting MHD equilibrium (originating from \textsc{Cliste} calculations \cite{Carthy12, Diarmuid+Schneider-priv}), the radial structure and frequency of the perturbation is calculated numerically with the linear gyrokinetic eigenvalue solver \textsc{Ligka} \cite{Lauber07}. \textsc{Hagis} is extended to read the real as well as the imaginary part of the electromagnetic perturbation. In general, the magnetic ($\tilde{\Psi} = (\nabla A)$) as well as the electric ($\tilde{\Phi}$) part can be taken into account separately. However, if the damping is small, as in this case, it can be assumed that $\tilde{\Psi} \propto \tilde{\Phi}$.\\
    During \textsc{aug} discharge \# 23824, two dominating modes are seen in the experiment \cite{Garcia10,Lauber09} at $t = 1.16$\ s around the radial positions $s \approx 0.3$ and $s \approx 0.5$ with frequencies of $120$\ kHz ($n=4$) and $55$\ kHz ($n=4$) respectively. At the later time point, the frequencies have evolved upwards, resulting in $160$\ kHz ($n=5$) and $70$\ kHz ($n=4$). The higher frequency mode is identified as TAE, the lower as RSAE (=AC) at $t = 1.16$\ s but as BAE at $t = 1.51$\ s \cite{Lauber09}.
     The following two sections are dedicated to the simulation of fast particle redistribution and losses under realistic simulation conditions with the vacuum extended version of \textsc{Hagis}. Therefore, the original eigenmodes given by \textsc{Ligka} are used with all their poloidal harmonics (unless otherwise noted), with a radial structure as depicted in \fref{ligka-input}a for the MHD equilibrium of \textsc{aug} discharge \#23824 at $t=1.16$\ s (inverted \qp) and in \fref{ligka-input}b for the time point $t=1.51$\ s (monotonic \qp). In the following, the two equilibria will be referred to as ``scenario1.16'' and ``scenario1.51'' respectively. The \textsc{Icrh}-like distribution function is used, and the fast particle beta value was chosen as \bef=0.02\%, a quite realistic value, that leads to mode amplitudes comparable to those measured experimentally \cite{Garcia10}, or slightly higher ($\delta B/B \in [5\cdot 10^{-3},2\cdot 10^{-2}]$).

  \subsection{Internal Transport Study in Two Different \qp\ Equilibria}
\begin{figure}[H]
      \centering
      \includegraphics[width=0.5\textwidth,height=5cm]{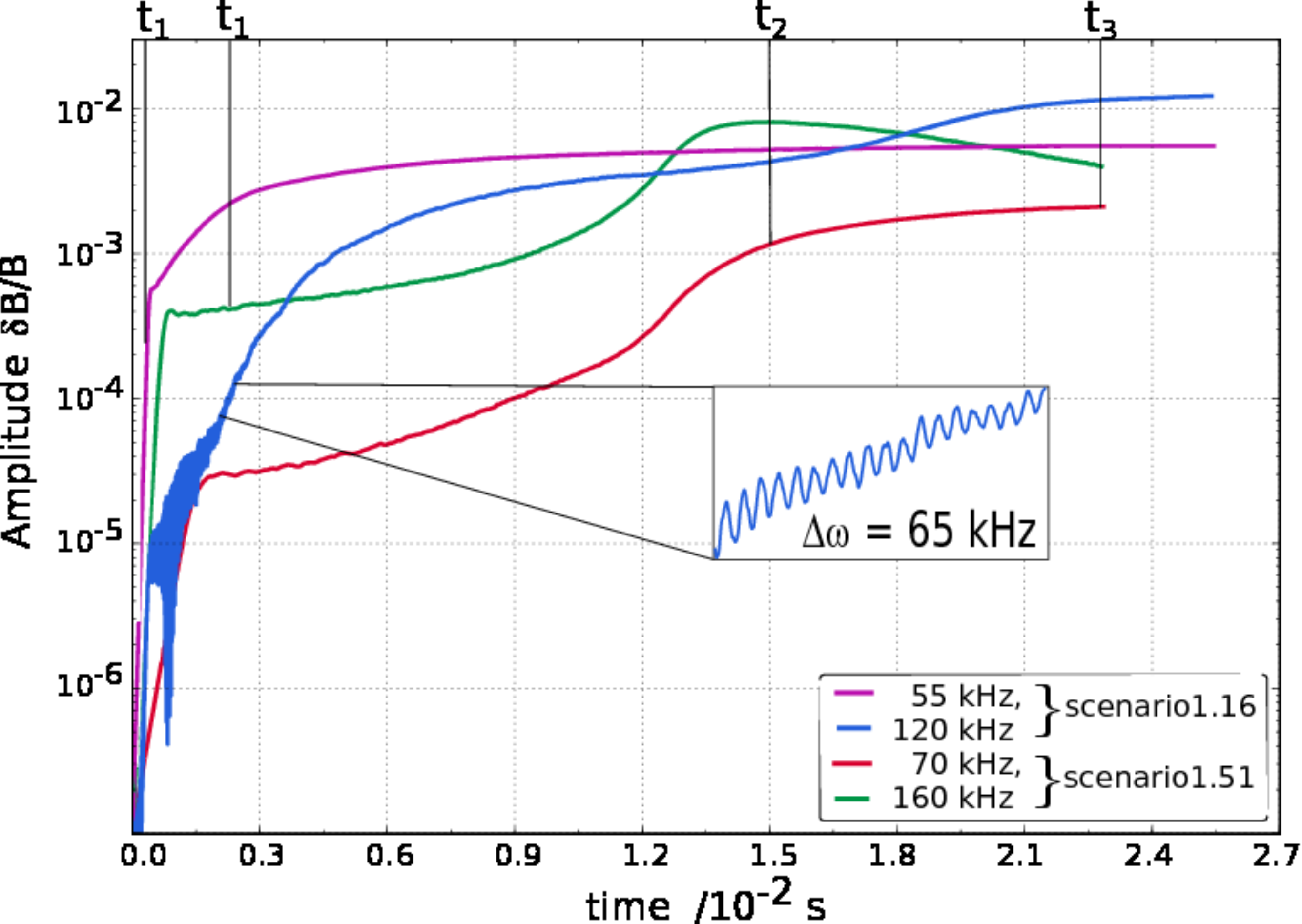} 
      \caption{\itshape Amplitude evolution in two double-mode simulations with different MHD equilibria 
        of the \textsc{aug} discharge \#23824 (\bef=0.02\%): pink and blue: $n=4$ RSAE ($55$\ kHz) 
        and $n=4$ TAE ($120$\ kHz) (radial structure see \fref{ligka-input}a), 
        simulated in the equilibrium at $t= 1.16$\ s, 
        red and green: $n=4$ BAE ($70$\ kHz) and $n=5$ TAE ($160$\ kHz) (see \fref{ligka-input}b) 
        in the equilibrium at $t= 1.51$\ s. 
        The marker are loaded according to the
        \textsc{Icrh}-like distribution function described in \sref{sec:simcond}.}
      \label{run0461+462_ampl}
    \end{figure}

     The mode amplitude evolution in both equilibria are shown in \fref{run0461+462_ampl}: in scenario1.16, one can see clearly the double-resonance effect as described in ref.\ \cite{Schneller12} leading a superimposed oscillation (insert), and also to the TAE (blue curve) exceeding the initially much faster growing RSAE (pink curve) in the late nonlinear phase. The examination below (\fref{run0461+462_Esdf-res}a for scenario1.16) confirms that both modes reach the stochasticity threshold. In scenario1.51, the TAE (green curve) grows much faster than the low frequency mode (in this case a BAE, red curve), but both grow slower and saturate at a lower amplitude compared to scenario1.16. The TAE's stochasticity threshold is reduced by the stochastization due to the BAE. Since the simulated time scale in this case is only one order of magnitude below the slowing-down time, the effect of energy dissipation becomes slightly visible with the TAE amplitude decreasing at the end of the simulation. Beyond this time point, the present model is not valid any more. One would have to extend it for a fast particle source term and take damping mechanisms into account.
    To understand the mode-particle interaction in the different stages of the mode evolution, it is helpful to look at the processes in phase space. In \fref{run0461+462_Esdf-res}b, redistribution in $E$-$s$ space is shown during the mode's resonant phase for scenario1.51. In this phase, the redistribution in phase space takes place along the resonance lines, from higher energies and lower radial positions to lower energies further outside. As the modes in scenario1.51 stay longer in the resonant phase, the redistribution pattern is better visible in this case, but can be seen in scenario1.16 as well.
    When the modes, or at least one of both, reach the stochastization level, a massive radial gradient depletion takes place over the whole energy space. No resonance pattern is visible any more, as visualized in \fref{run0461+462_Esdf-res}a (for scenario1.16). This happens in both scenarios, however much later in scenario1.51 around $t \approx 1.5 \cdot 10^{-2}$\ s whereas at $t \approx 1.5\cdot 10^{-3}$\ s in scenario1.16. In both cases, at lower energies ($E \in [50,200]$\ keV), the resonance regions are still slightly visible in the redistribution pattern. However, only in scenario1.16 does the redistribution (along the $p = 0,~1$ resonance lines of both modes) cross the loss boundary. The redistribution takes place independently on the energy, but once the radial gradient is depleted in the energy region of $E \in [400,600]$\ keV, the redistribution is radially broadened at lower energies. In \sref{sec-poloidalharmonics}, it will be shown that this broad radial redistribution is caused by the outer poloidal harmonics of the TAE.

  \subsection{Fast Particle Losses in Two Different \qp\ Equilibria}
\begin{figure}[H]
      \centering
      \subfigure[\itshape scenario1.16, $t_2=1.5\cdot10^{-2}$\ s.]{\includegraphics[width=0.4\textwidth, height=5.5cm]{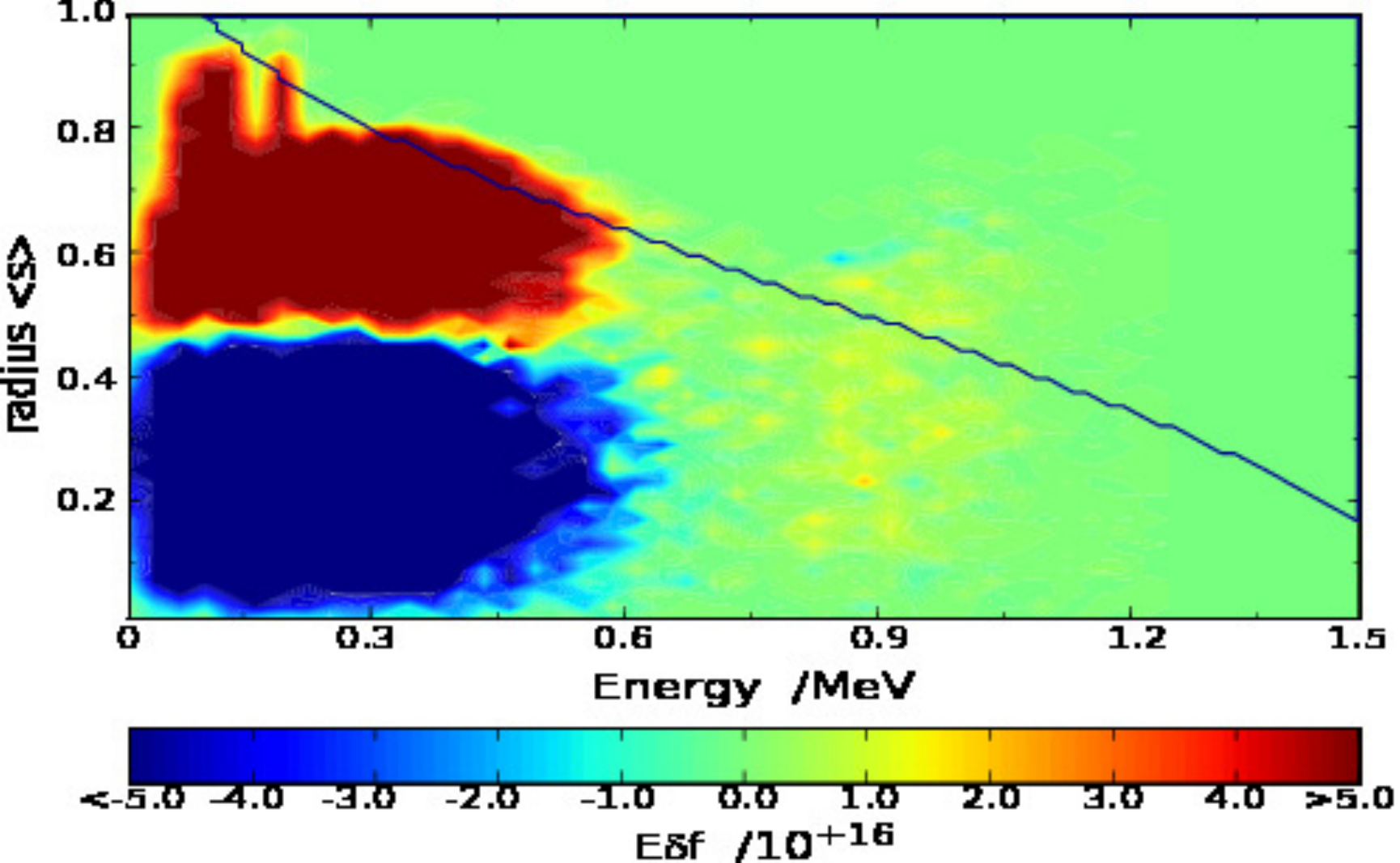}} 
      \hspace{0.7cm}\subfigure[\itshape scenario1.51, $t_1=2.3\cdot10^{-3}$\ s.]{\includegraphics[width=0.4\textwidth, height=5.5cm]{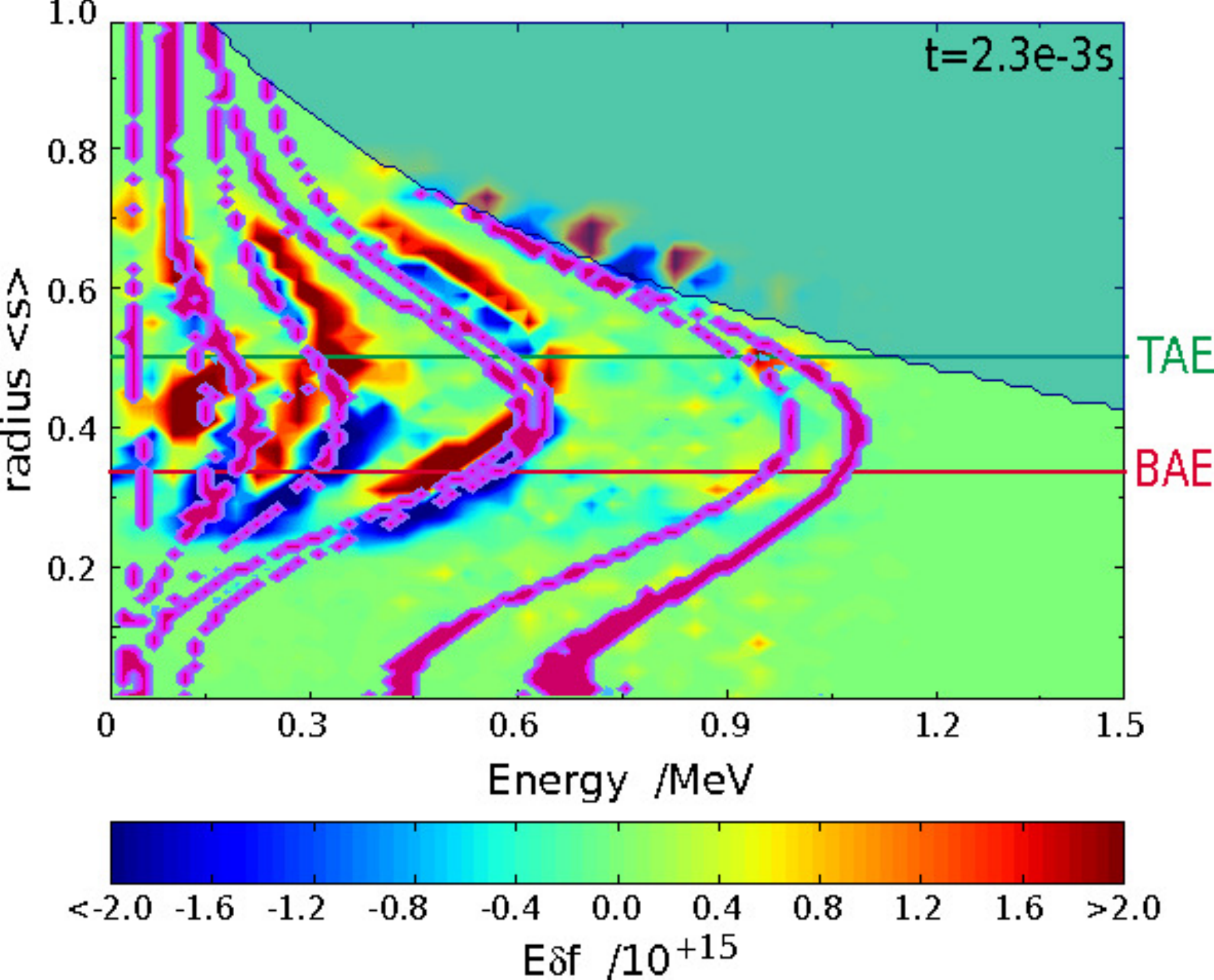}} 
      \caption{\itshape Redistribution (in terms of fast particle pressure $E \delta f$) in energy and radial space at two different stages in the simulation. Red indicates particle accumulation, blue means particles move away. In b, the pink lines give the resonance lines for trapped particles, the blue areas the loss region (both according to \fref{resplot_AUG-23824}b). The horizontal lines denote the modes' radial positions. One can see clearly the redistribution caused by the wave-particle resonance along the resonance lines in the mode regions.}
      \label{run0461+462_Esdf-res}
    \end{figure}

\begin{figure}[H]
      \centering
      \subfigure[\itshape scenario1.16]{\includegraphics[width=0.48\textwidth,height=5.5cm]{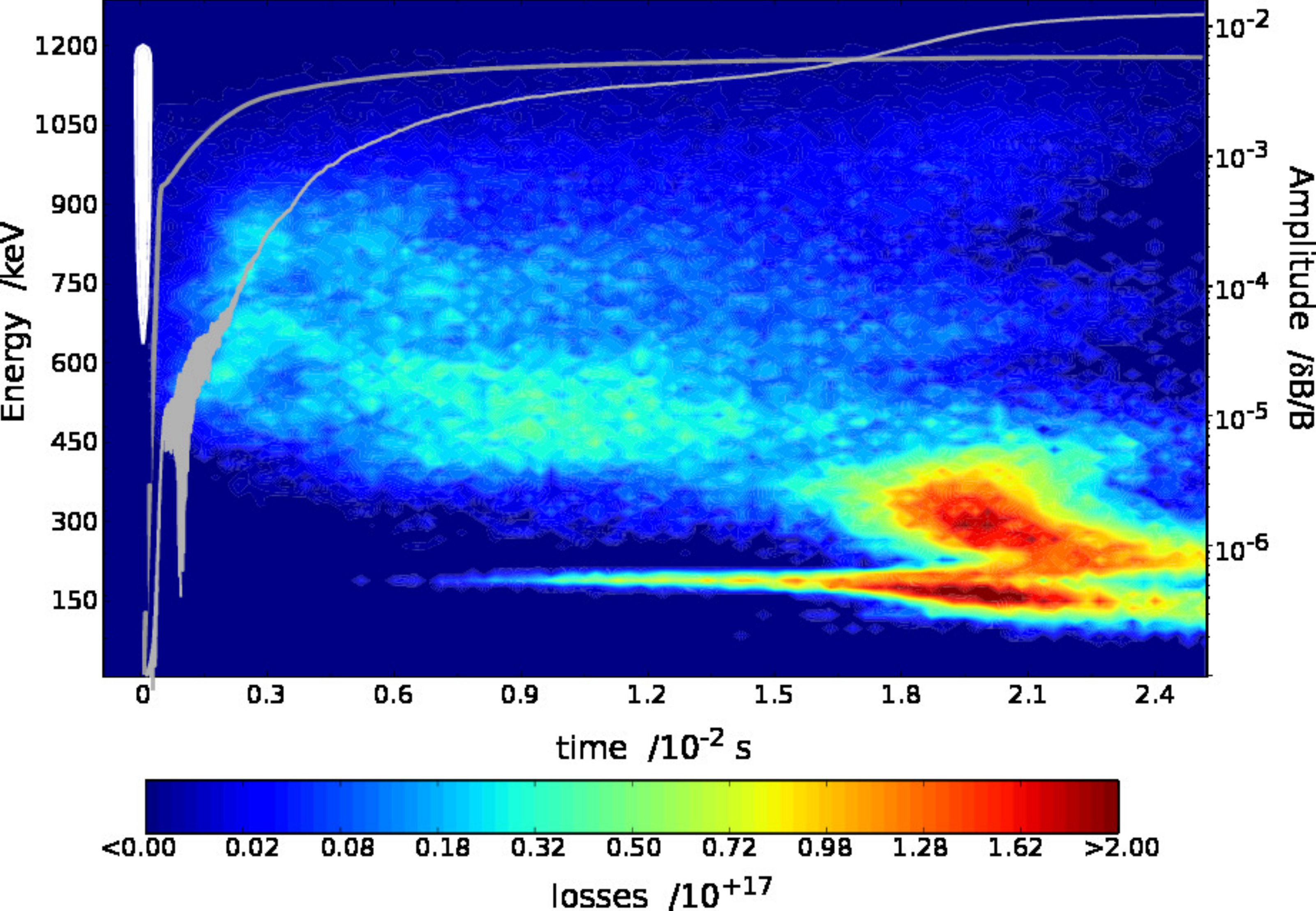}} 
      \hspace{0.3cm}\subfigure[\itshape scenario1.51]{\includegraphics[width=0.48\textwidth,height=5.5cm]{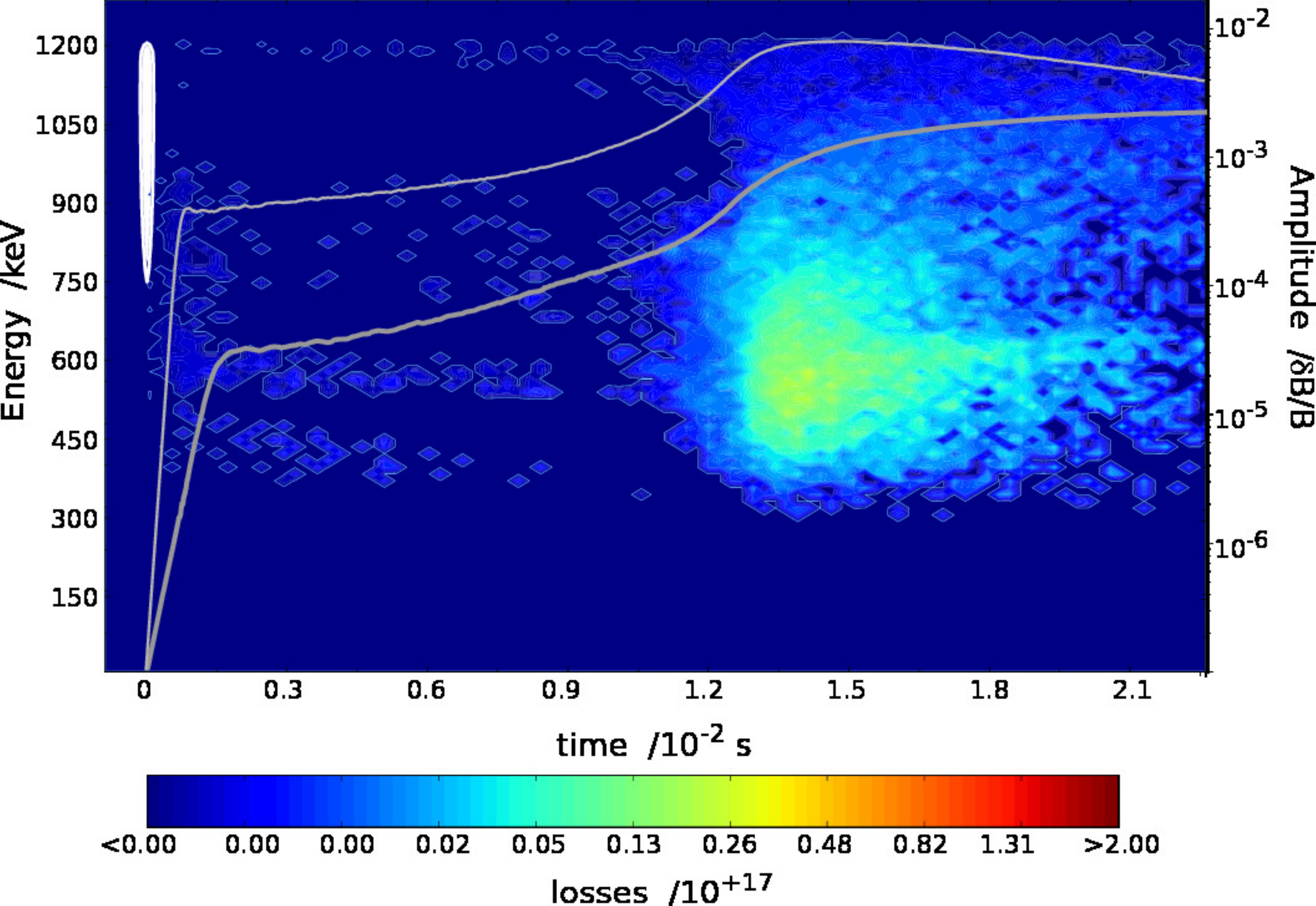}} 
      \caption{\itshape Temporal evolution of losses appearance at the first wall in energy space for the
      two different scenarios. The white areas indicate the prompt losses as carried out in an extra 
      simulation with a thinner $\Lambda$ distribution function \eref{formula:lampollim2}, 
      as explained before. The gray lines give the mode amplitude evolution as shown
      in detail in \fref{run0461+462_ampl}.} 
      \label{run0461+462_tEloss}
    \end{figure}

    The redistribution plots already indicate fewer losses in scenario1.51, due to the smaller loss area, and the weaker redistribution especially in the lower energy range. In fact, scenario1.16 gives much more losses --  prompt (i.e.\ appearing before any influence of the mode) as well as losses in the later phase of the simulation (compare \fref{run0461+462_tEloss}a and b). In scenario1.51, there are no losses with an energy below $300$\ keV at all. This is in accordance with the redistribution pattern of this case, giving no redistribution across the loss boundary for $E < 300$\ keV, due to the smaller loss region. The prompt losses appear in both cases in the higher energy range. They reach down to $600$\ keV (scenario1.16) and $750$\ keV (scenario1.51).\\
    As the resonant phase is not distinguishable very clearly in scenario1.16, it is difficult to find resonant losses at very distinct energies. However, the losses in this phase have energies that match with the resonant phase space redistribution across the loss boundary, although not as clearly as in scenario1.51 (see \fref{run0461+462_Esdf-res}b).  In scenario1.16, the resonances are at energies of $E \in [450,700]$\ keV and $\in [750,950]$\ keV. In scenario1.51, the distinct energies result in losses around $400$\ keV, $550$\ keV, $700$\ keV and $850$\ keV (see \fref{run0461+462_tEloss}b).
    When stochastization sets in, a higher amount of energetic losses appears, due to the broad redistribution across the loss boundary over a large energy range. In scenario1.16, the losses reach down to low energies. The cause is a phase space channeling effect within the redistribution (according to the domino effect proposed theoretically in \cite{Berk95-III}), as discussed above (\fref{run0461+462_Esdf-res}a): first, energies $E \in [300,600]$\ keV are redistributed until the radial gradient in this energy area has flattened and the redistribution continues in $E \in[200,400]$\ keV. In the loss spectrum over time (\fref{run0461+462_tEloss}), the successive transport across the loss boundary can be recognized as losses. Due to the many poloidal harmonics resulting in a very broad radial structure of the TAE (as will be shown in \sref{sec-poloidalharmonics}), even a very low energy resonance is able to transport particles across the loss boundary, that in turn appear as a thin loss peak at $E \in [100,150]$\ keV, once the TAE has reached the stochastization level.

  \subsubsection{Influence of the $q$ Profile}
\begin{figure}[H] 
      \centering
      \includegraphics[width=0.7\textwidth,height=4cm]{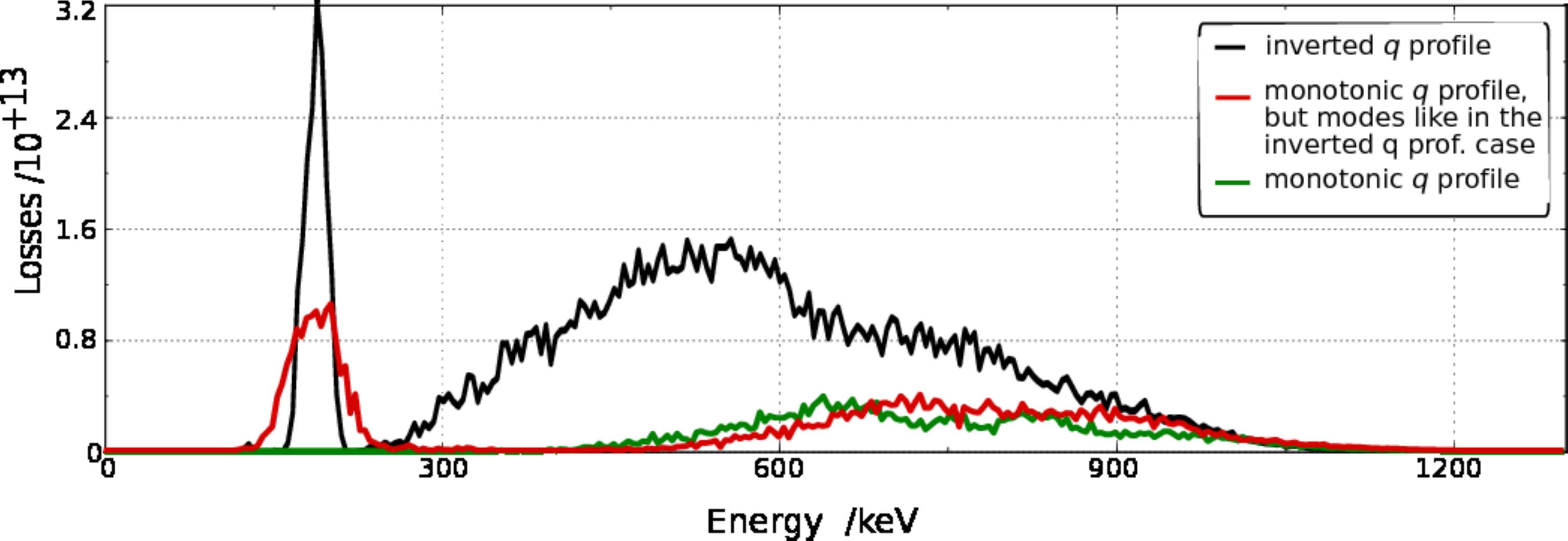} 
      \caption{\itshape Energy spectra of losses as numerically calculated within scenario1.16 (black curve),
        within scenario1.51 (green curve) and as simulated within the monotonic \qp\ equilibrium, 
        using the eigenmodes given by \textsc{ligka} for the inverted \qp\ equilibrium 
        (\fref{ligka-input}a) (red curve). 
        The amplitudes for these simulations were fixed at $\delta B/B = 5.1\cdot 10^{-3}$. 
        The losses shown appeared in a time interval starting 
        after c.a.\ 10 RSAE wave periods ($t \in [0.2,1.5] \cdot 10^{-3}$\ s).} 
      \label{run0502+506+507_Eloss}
    \end{figure}

    In scenario1.51, where the loss region is smaller, the TAE is not broad enough, to redistribute particles with $E < 350$\ keV across the loss boundary. Although one reason for this is the lower amplitude levels reached in scenario1.51 (compared to scenario1.16), the main reason for the missing of the lowest energetic losses (around $\approx 180$\ keV) and the small amount in the whole range of $E \in [300,600]$\ keV is found in the different equilibrium, i.e.\ the smaller loss region and the larger distance between the resonance lines. This is proved by simulating the same TAE and RSAE eigenmode structures of scenario1.16 in both equilibria with a fixed mode amplitude of $\delta B/B = 5.1\cdot 10^{-3}$ (an experimentally realistic value): as shown in \fref{run0502+506+507_Eloss}, even with the same mode amplitude, structure, mode number $n$ and frequency, the monotonic \qp\ case leads to far fewer losses. No losses at all are observed in $E \in [300,600]$\ keV. The complete inhibition of the very low energy peak around 180\ keV however, is caused by a combination of the monotonic \qp\ equilibrium and the corresponding eigenmode.\\
    The drop in the height of the very low energy peak not only results from the smaller loss area in the monotonic \qp\footnote{$~$Note that the resonances for the scenario resulting in the red curve differ from what is shown in \fref{resplot_AUG-23824}b, because as mode frequencies, the values from the modes in the inverted \qp\ equilibrium were used.}, but also from a different redistribution mechanism: due to the less dense resonances in phase space around $E \in [100;600]$\ keV, the redistribution in the monotonic \qp\ is less efficient but spreads wider towards higher energies. In combination with the different loss boundaries, this leads to the missing of losses in the energy range of $E \in [300,600]$\ keV. The higher the energy is, the smaller is the effect of the equilibrium on the loss spectrum. This is quite logical, since high energy particles are on large orbits, and therefore require only a small perturbation to be expelled.

  \subsubsection{Influence of the Poloidal Harmonics}\label{sec-poloidalharmonics}
    Next, the effect resulting from the many poloidal harmonics of the TAE is investigated. By calculating the eigenmodes with the \textsc{Ligka} solver, the poloidal harmonics are known and can be used within the \textsc{Hagis} simulation. Comparing two simulations -- one with the poloidal harmonics $m=4$ and $m=5$ only, the other one with all harmonics (from $m=4$ to $m=10$) as shown in \fref{ligka-input}a (lower) -- shows a similar qualitative mode amplitude evolution with similar mode saturation levels. However, the TAE growth rate is lower if simulating with all poloidal harmonics, as the energy has to be distributed over more poloidal harmonics. In the saturation phase, this effect is compensated by the wider radial range, i.e.\ the mode with all poloidal harmonics is able to tap energy from the radial gradient also radially further outside.
     \begin{figure}[H]
      \centering
      \subfigure[\itshape Redistribution in phase space ($t =3.6 \cdot 10^{-3}$\ s)]{\includegraphics[width=0.48\textwidth,height=5.5cm]{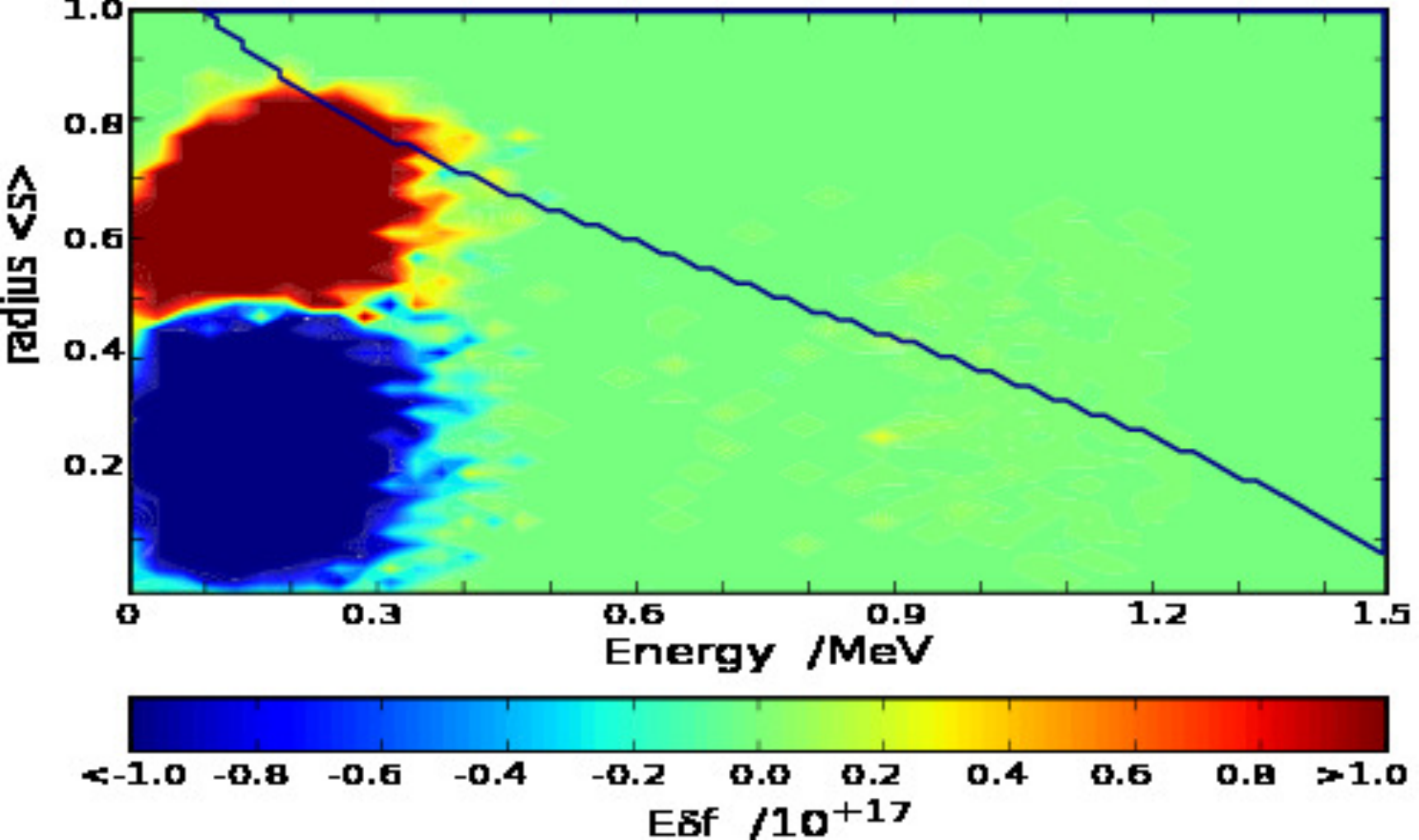}} 
      \hspace{0.3cm}\subfigure[\itshape Losses over time]{\includegraphics[width=0.48\textwidth,height=5.5cm]{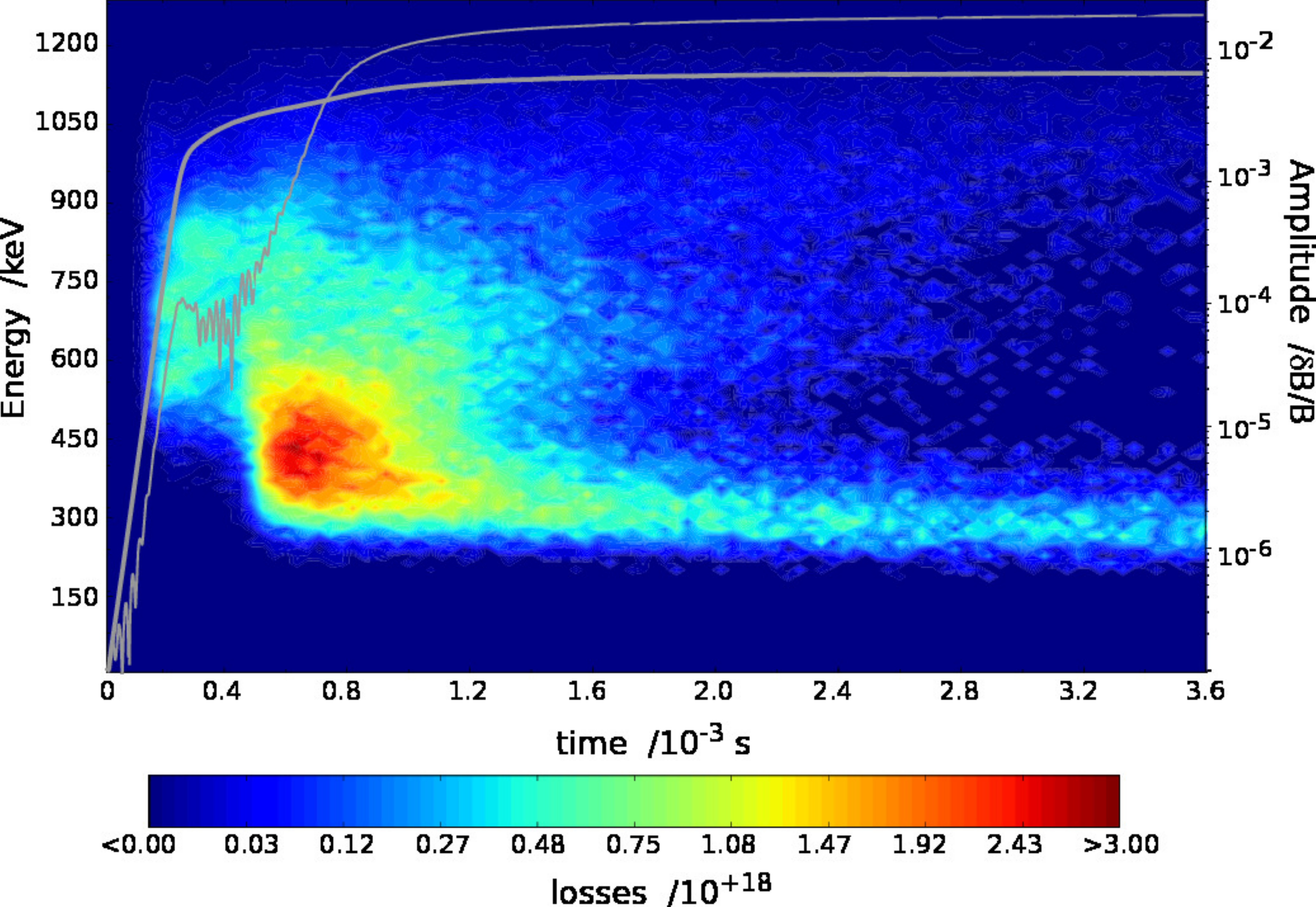}} 
      \caption{\itshape Redistribution (a, blue means particles move away, red means they accumulate)
        in energy and radial position space and losses at the first wall (b) of a simulation  
        in the inverted \qp\ (\bef=0.05\%), with modes given by \textsc{Ligka} (see \fref{ligka-input}a),
        but the TAE is simulated with only its first two poloidal harmonics. 
        The blue line in a gives the loss boundary (as explained in \fref{resplot_AUG-23824}a), the gray
        curves in b the mode amplitude evolution.}
      \label{run0451_Esdf+tEloss}
    \end{figure}
    This effect is well visible in the redistribution: only if all poloidal harmonics are present, the TAE redistributes particles at very low energies (around $E \approx 100$\ keV) and far outside radial positions ($s > 0.8$), as becomes clear when comparing \fref{run0461+462_Esdf-res}a with \fref{run0451_Esdf+tEloss}a. This redistribution results in the observed low energy losses of $E < 200$\ keV, that are not observed in the simulation with only two poloidal harmonics (compare \fref{run0461+462_tEloss}a and \fref{run0451_Esdf+tEloss}b).\\
    In summary, it is clear now, that the losses with $E < 200$\ keV result from the combination of the broad TAE with many harmonics and the large loss region appearing in the inverted \qp\ equilibrium. However, it was shown above (see \fref{run0502+506+507_Eloss}), that significantly less of these losses appear when simulating the same modes at the same amplitudes in the monotonic \qp, due to the smaller loss region. Concerning the losses in the lower energy range between around 300\ keV up to c.a.\ 600\ keV, it was found that these are due to the dense resonance lines meeting the loss boundary in the inverted \qp\ scenario, quite independent of the mode structure. In the monotonic \qp, these losses do therefore not occur.

  \subsubsection{Double Mode Effect}
      As the mode amplitudes in the nonlinear saturation are known now from simulations with consistent mode evolution, \textsc{Hagis} is run with these mode amplitudes kept fixed. This offers the possibility to investigate the losses caused by each of the modes alone vs.\ the losses caused by double-mode interaction, but without the additional effect that different amplitudes are reached in single-mode simulations compared to double-mode simulations. \Fref{run0502+503+504_Eloss} shows the energy spectra of the losses appearing in such double-mode simulation (black curve), as well as those obtained, when simulating both modes individually (blue for the TAE and pink for the RSAE). Note that the loss spectra curves obtained in the single-mode simulations are shown multiplied by a factor of 10. For each of the three spectra, the same time interval is chosen, starting after approximately 10 wave RSAE periods, to avoid the effect of prompt losses. All mode amplitudes are fixed at $\delta B/B = 5.1\cdot 10^{-3}$, which is a reasonable value, when comparing to experimentally measured amplitudes. Further, it is of the order of the minimum saturation level reached in the previous simulations with consistent mode evolution. It is not surprising that the more core-localized RSAE causes significantly fewer losses than the TAE, which is broad and located at a higher radial position. However, for both single-mode simulations, the losses' energy spectra exhibits clearly the different resonances: losses appear at energies, where resonance lines meet the loss boundary (with \fref{resplot_AUG-23824}a the bounce harmonic numbers $p$ can be identified). But the most interesting fact is found, when looking at the double-mode simulation's losses: they do by far exceed the sum of the losses of the single-mode simulations, although the amplitude levels are the same. The redistribution in phase space is consistent with the appearance of losses in the double-mode simulation versus the single-mode simulations. As well as the losses, also the redistribution in the double-mode simulation exceeds by far those of the single-mode cases. Thus, the large number of losses in multi-mode situations is not only caused by the higher mode amplitudes that are possibly reached then (as was shown in ref.\ \cite{Schneller12}). The same reason that leads to higher mode amplitude levels -- the lowering of the stochasticity threshold -- also causes more losses, because the redistribution is enhanced by the stochastization. If the loss boundary is large, as in the inverted \qp\ equilibrium, this redistribution leads to a large number of losses.\\
    \begin{minipage}{0.48\textwidth}
    \vspace{2cm}
    \begin{figure}[H]
      \centering
      \includegraphics[width=1\textwidth,height=3.5cm]{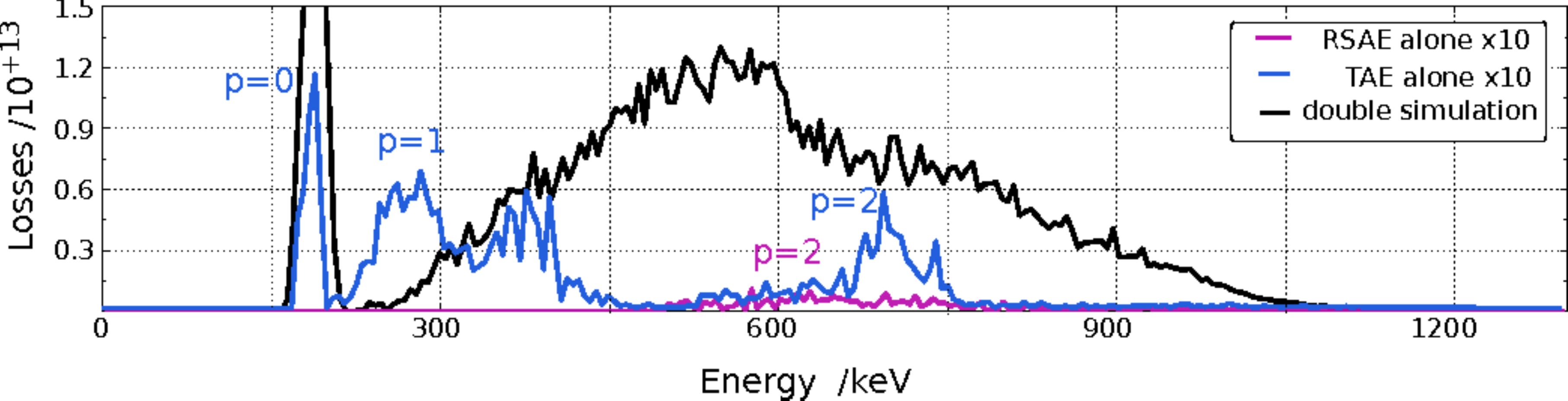} 
      \caption{\itshape Energy spectra of losses in three different}
      \label{run0502+503+504_Eloss}
    \end{figure}
        \vspace{-0.5cm}{\itshape \small simulations with fixed mode 
        amplitudes at $\delta B/B = 5.1\cdot 10^{-3}$ for a time interval starting after ap\-prox\-imately 
        10 RSAE wave periods ($t \in [2.0,1.2] \cdot 10^{-3}$\ s): the pink and the blue curves give
        the loss spectrum of the RSAE and TAE single-mode simulation, stretched by a factor of 10. 
        The black curve gives the losses appearing if simulating both modes (\fref{ligka-input}a) together.} 
        \vspace{0.4cm}
    \end{minipage}
    \hfill\begin{minipage}{0.50\textwidth}
    \begin{figure}[H]
      \centering
      \includegraphics[width=1\textwidth,height=5.5cm]{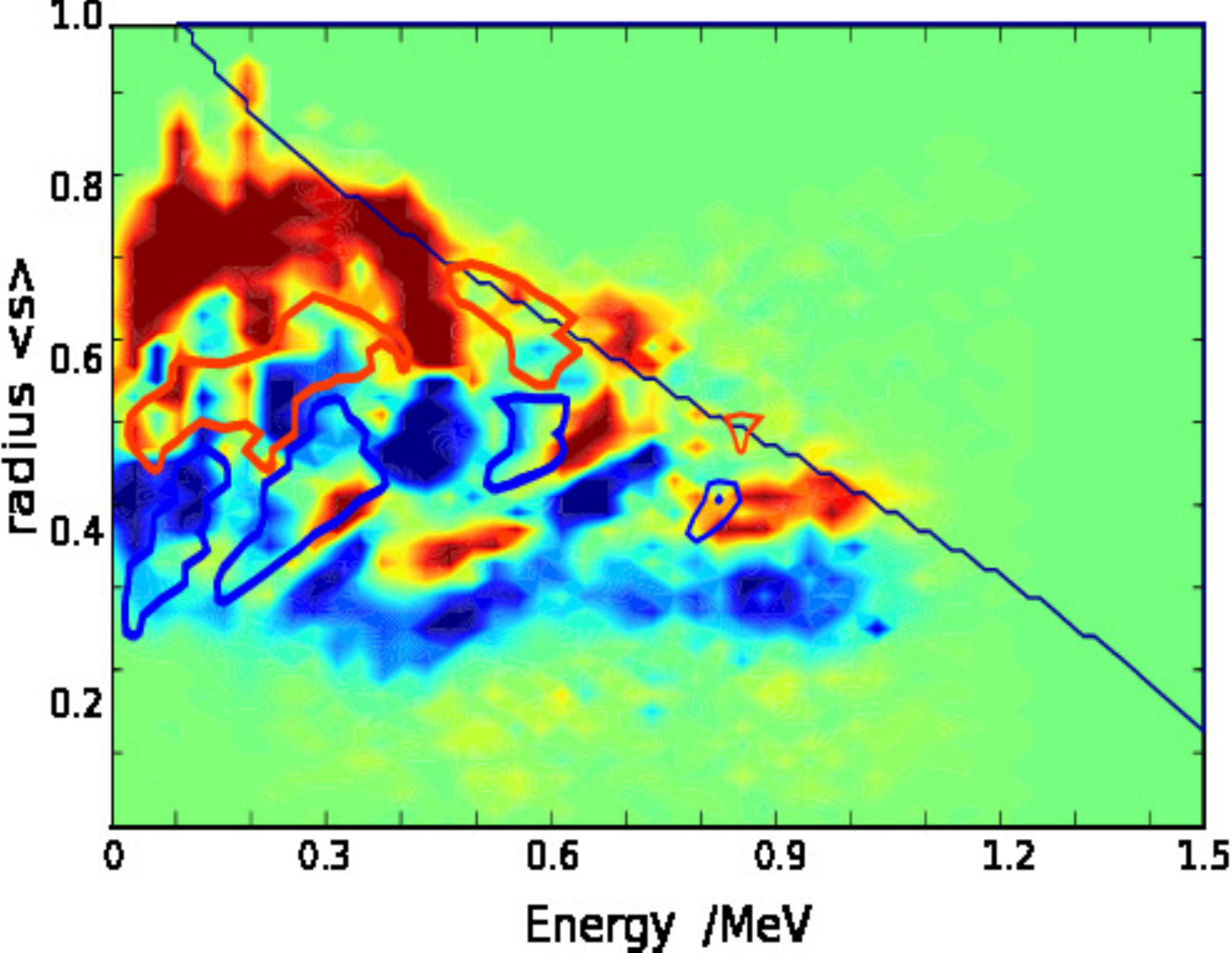} 
      \caption{\itshape Redistribution in energy and radial position} 
      \label{run0503+504_Esdf3300}
    \end{figure}
        \vspace{-0.5cm}{\itshape \small space caused by the TAE (color code) and 
        the RSAE (boundary lines) in the fixed amplitude simulation (at $\delta B/B = 5.1\cdot 10^{-3}$). 
        Red indicates particle ac\-cu\-mu\-la\-tion (higher $E\delta f$), blue that 
        particles move away (lower $E\delta f$). 
        The dark line gives the loss boundary of the inverted \qp\ equilibrium.}
      \vfill
    \end{minipage}\\
    In the double-mode scenario, stochastization is enhanced due to the dense coverage of phase space with both resonance lines and perturbation structure. In \fref{run0503+504_Esdf3300}, this effect can be observed clearly: the RSAE transports energetic particles into phase space areas from where they are further redistributed by the TAE. Since this effect also occurs vice versa, both modes mutually refill the phase space areas with particles, where the other mode redistributes them away from.

  \subsubsection{Comparison of the Loss Ejection Signature}
      The first striking statement of ref.\ \cite{Garcia10} is the difference in the \textbf{total amount of losses} between the earlier time points, $t < 1.4$\ s, when a large number of losses appears in the whole energy range, and later times, $t > 1.4$\ s, when there are only few losses in the high energy channel \cite{Garcia10} and almost none in the low energy channel \cite{GarciaIAEA09}. 
    Focusing on the two time points $t=1.16$\ s and $t=1.51$\ s within the experimentally measured loss signal reveals (figure 4 of ref.\ \cite{Garcia10}): at $t=1.16$\ s, an equally high amount of losses is found in the low and the high energy channel, whereas at $t=1.51$\ s, no low energy losses appear -- except for a general noise level, and fewer ($\approx 18\%$) in the high energy range. This result is well reproduced by the numerical simulations, where scenario1.16 gives a large number of losses in the lower and in the higher energy range, with quite comparable levels, relative to each other. In scenario1.51, no losses appear below 300\ keV  and a few for higher energies ($E \gtrsim 300$\ keV), mostly within $E \in [550;750]$\ keV. The drop of losses in the higher energy range between scenario1.16 and scenario1.51 is in the range of around 10\% in the simulation. However, to compare the losses quantitatively with the ones measured at the \textsc{Fild} is only possible in the lower energy channel, since in the higher energy channel, prompt losses are superimposed. These prompt losses cannot be quantified within the model, as explained in \sref{sec:simcond}.\\

\begin{figure}[H] 
      \centering
      \subfigure[\itshape {Losses' signature for the energy range of $E\in [90,350]$\ keV. Light blue: underlying incoherent losses.}]{\includegraphics[width=0.8\textwidth,height=3cm]{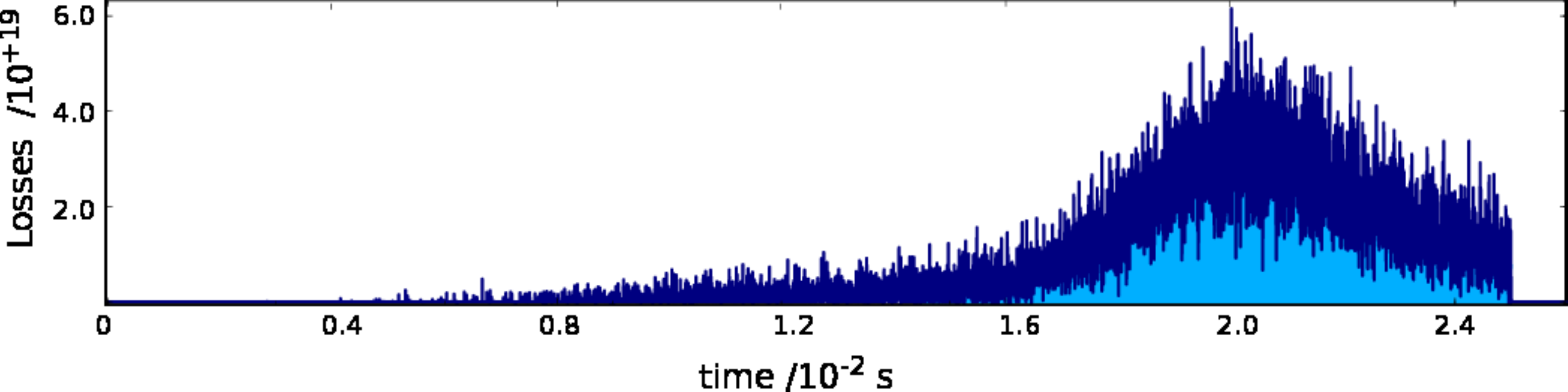}} 
      \subfigure[\itshape {Time integrated losses with $E\in [90,350]$\ keV over TAE amplitude (green) and quadratic fit (red).}]{\includegraphics[width=0.8\textwidth,height=3cm]{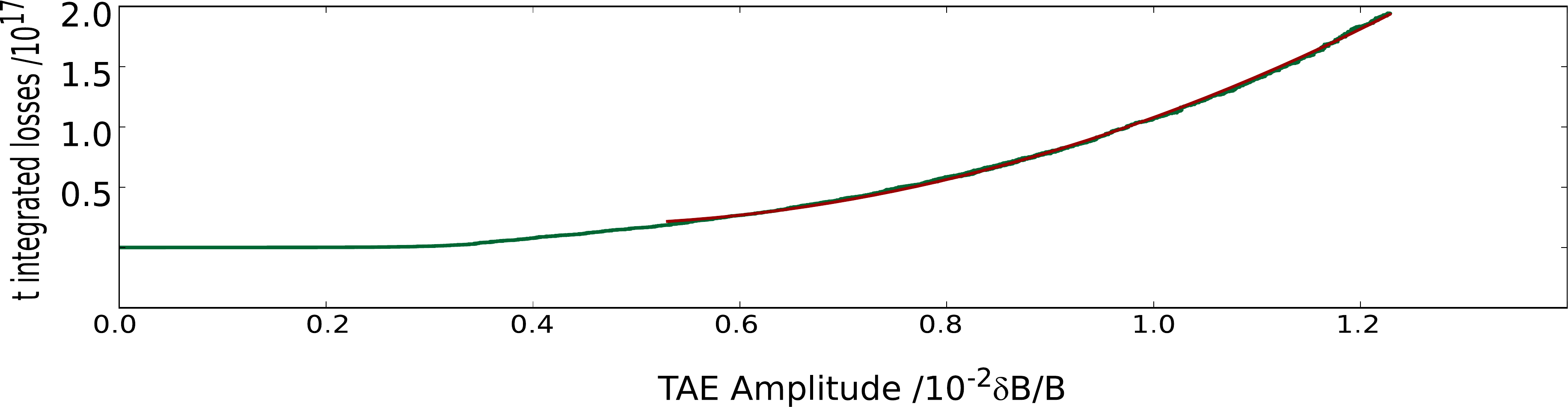}} 
      \caption{\itshape Losses at the first wall in scenario1.16.}
      \label{run0461_tloss}
    \end{figure}

     Next, the \textbf{nature of losses -- incoherent or coherent} is discussed. In the experiment, incoherent diffusive losses are observed \cite{Garcia10}, identified by a typical quadratic scaling with the mode amplitude  ($\propto (\delta B/B)^2$) \cite{Sigmar92}. It is an important result of the presented realistic simulations, that this quadratic scaling is now found also numerically (\fref{run0461_tloss}b). Further, the quadratic scaling is another evidence for the diffusive character of the losses in the lower energy range of scenario1.16. The simulations for $t=1.51$\ s (see \fref{run0461+462_tEloss}b) give losses only in the high energy range. There are both prompt, i.e.\ incoherent losses and non-prompt losses, which show clearly an energy spectrum correlated to the resonance energies at the loss boundary. However, there is also a small amount of non-prompt incoherent losses (diffusive). Thus, the few incoherent losses seen also in the \textsc{Fild} signal can be both prompt or diffusive. But taking into account, that the mode amplitudes were slightly overestimated in the simulation, it can be concluded that there are only very few losses with diffusive origin in the experiment.
     In the experiment, the ratio of coherent to incoherent losses at $t=1.16$\ s is about 1:5 in the low and roughly 1:2.5 in the high energy channel. A possible overestimation of incoherent losses due to the smearing effects (caused by a small deuterium fast ion population, magnetic field ripple, etc.) has to be taken into account. In the simulation, coherent (resonant) losses are found as well, especially in the medium to higher energy range. They can be clearly identified via a Fourier analysis of the ejection signature, which revealed the main peak at the mode's beat frequency. The maximum ratio of incoherent (caused by phase space stochastization, combined with the large loss region of the inverted \qp) to coherent (resonant) losses appears in the low energy range and is about 1:1 (as illustrated by \fref{run0461_tloss}a). Thus, it is lower than the experimental finding, but might be increased when taking into account more than two modes. In the higher energy range, a quantitative analysis is not possible due to the prompt losses (as mentioned above). Though, a rough estimate is, that a large fraction of the losses in this energy range is prompt: since no prompt losses occur in the low energy range, one can calculate a scaling factor between the experimental loss amplitude and the simulated one. With this factor, the simulated losses in the high energy range can be compared to the measured ones. The difference gives the prompt losses, that are missing in the simulation. In the case of the most realistic simulation, this allows one to estimate, that between around 5\% and 50\% of the measured incoherent losses in the high energy channel would be prompt. However, this can be only a rough estimate, due to the uncertainties that enter the comparison. Besides the limitations of the model (discussion see ref.\ \cite{mirjam_phd}), these are the width of the energy channels, the \textsc{Fild} noise level, and the fact, that the scaling between experiment and simulation can depend on the energy, since the losses at different energies are caused by different loss mechanisms. If one considers smearing effects at the \textsc{Fild}, that overestimate the incoherent losses, the amount of prompt losses within the incoherent losses are expected to be higher.

  \subsubsection{Comparison of the Phase Space Pattern} 
\begin{figure}[H]
        \centering
        \includegraphics[width=0.7\textwidth,height=5.5cm]{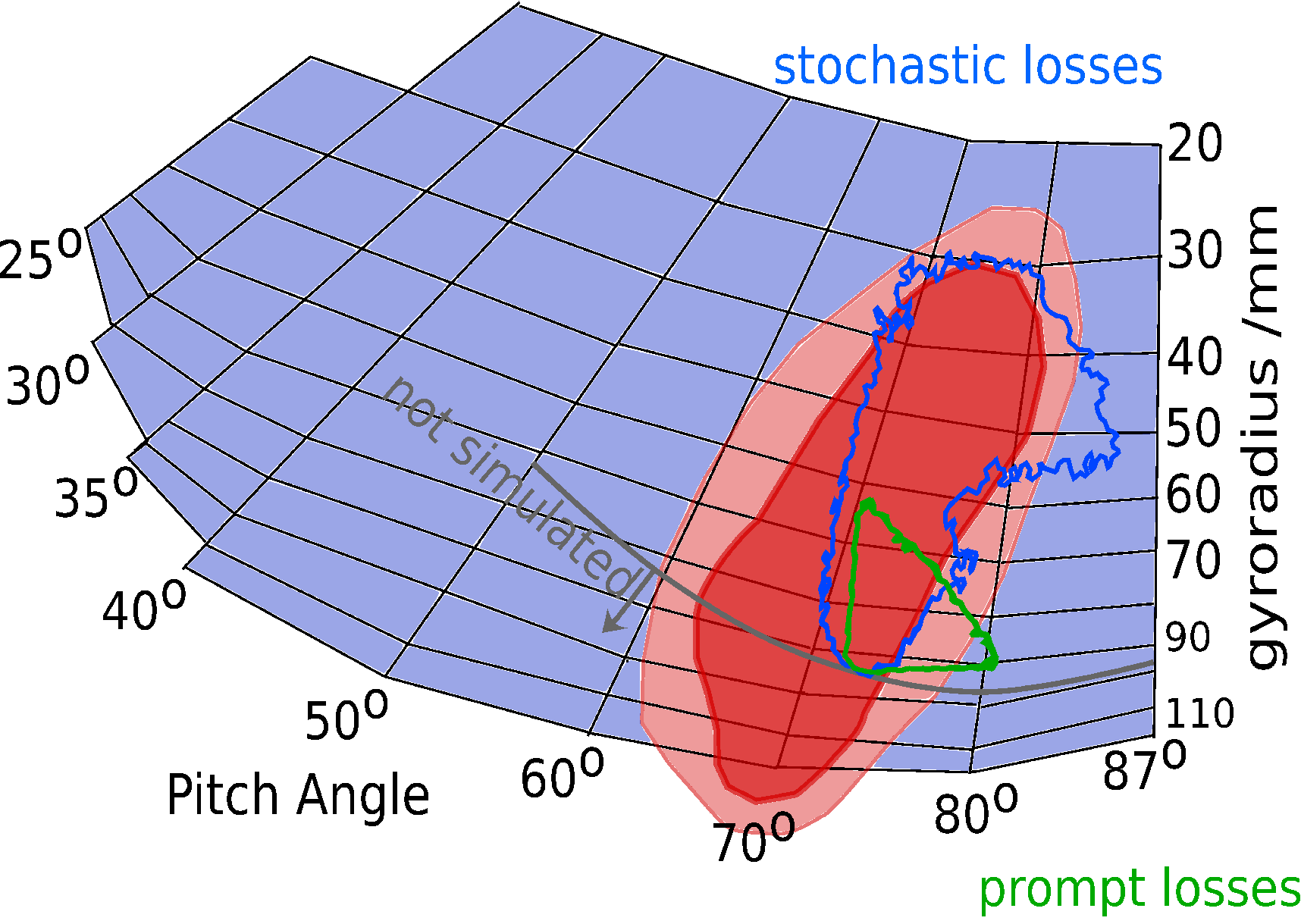} 
        \caption{\itshape Loss pattern (colored red) in phase space 
          (gyroradius $\gyrad$ over pitch angle $\pitch^o$) as 
        measured at the \textsc{Fild} in \textsc{aug} discharge \#23824 at time $t \in [1.14,1.16]$\ s 
        from ref.\ \cite{GarciaIAEA09}.
        Further, the loss pattern as resulting from the simulation of an \textsc{Icrh}-generated 
        fast particle distribution function in the MHD equilibrium of \textsc{aug} \#23824 at $t= 1.16$\ s 
        with the corresponding perturbation 
        given by \textsc{Ligka} (see \fref{ligka-input}a) is shown.
        For the distribution function see \sref{sec:simcond}, 
        esp.\ \eref{formula:Lambdis},\eref{formula:lampollim}.
        The blue line gives the boundary of the non-prompt losses at the first wall -- where the majority 
        appears in the respective lower $\gyrad$ part within the boundary. 
        The green line depicts the boundary of the prompt losses' appearance at the
        first wall (simulated with the distribution function 
        of \eref{formula:Lambdis},\eref{formula:lampollim2}). Since the simulation range in the energy was restricted to $E \leq 1.2$\ MeV,
        there is an area in $\gyrad$-$\pitch^o$ space that has not been simulated: it is indicated by the gray line.
        The numerical values have been drift-corrected as explained in ref.\ \cite{mwb_phd}.}
        \label{Numeric_vs_Exp-116}
      \end{figure}

      In the next step, the \textbf{energy-pitch angle characteristics }of the numerical losses is compared with the experimental measurement. Using the synthetic \textsc{Fild} diagnostics as described in ref.\ \cite{mwb_phd}, it is found that experimental (colored red in \fref{Numeric_vs_Exp-116}, from ref.\ \cite{Garcia10}) and numerical loss pattern (green and blue line for the boundary of the prompt and the non-prompt loss pattern) match very well in phase space. The numerical results indicate, that the higher energy (or gyroradius-) incoherent losses are prompt losses, whereas the lower energy (or gyroradius-) incoherent losses are caused by phase space stochastization due to the presence of multiple modes. The highest energies (gyro radii) detected at the \textsc{Fild} do not appear in the simulation, as the maximum energy simulated was $E_\mathrm{max} = 1.2$\ MeV (indicated by the gray line in \fref{Numeric_vs_Exp-116}).\\
      To allow for a direct comparison to the \textsc{Fild} measurement, the numerical values in \fref{Numeric_vs_Exp-116} have been drift-corrected. This means that both energy (gyro radius) and pitch of each lost particle has to be shifted to account for a deviation caused by the experimental method of the \textsc{Fild}: since the particle continues drifting within the detector, the measured gyro radius and the pitch angle are overestimated. Ref.\ \cite{mwb_phd}, p.\ 115-117 gives a more detailed explanation and calculated the deviation for comparable cases to $\approx 6$\% in the gyro radius and $\approx 9^o$ in pitch angle.

\section{Conclusions and Outlook}
    In this work, multi-mode Alfv{\'e}nic fast particle transport was modeled with the vacuum-extended \textsc{Hagis} code \cite{mwb_phd}. One aim of the investigation was to compare numerical fast particle losses with experimental measurements of the \textsc{Icrh}-heated \textsc{Asdex} Upgrade discharge \#23824 \cite{Garcia10} and to gain a deeper understanding of the transport processes. This was achieved through the investigation of the internal redistribution in combination with existing resonances, as well as phase space and frequency analysis of the losses. The simulations have been carried out within a realistic model in various respects: on the energetic particle side, a more general, consistent \textsc{Icrh}-like distribution function was implemented. It accounts for the strong anisotropy of the \textsc{Icrh}-generated fast particle population. However, it is still assumed to be separable in its coordinates. This is a constraint concerning a realistic representation that is hoped to be overcome soon, especially for the radial and energy coordinates: the analytical model from ref.\ \cite{Troia12}, adjusted to realistic conditions with the help of the \textsc{Toric-ssfpql} code is planned to be implemented into \textsc{Hagis}. For the results presented in this work, crucial changes were ruled out via a sensitivity scan on the distribution function.

    On the perturbation side, \textsc{Hagis} has been extended to read kinetic perturbation data from the eigenvalue solver \textsc{Ligka} \cite{Lauber07}. The simulated losses' phase space pattern was found to coincide very well with the experimental one. Especially in multi-mode scenarios with different mode frequencies, stochastic redistribution sets in over a broad energy range, leading to lower energetic diffusive (incoherent) losses. The quadratical scaling of the amount of losses with the mode amplitude, which is predicted by theory \cite{Sigmar92} and seen in the experiment could be reproduced in realistic numerical simulations. Resonant losses appear from the late linear phase on, mainly in the intermediate to higher energy range, showing good coherence with the mode frequencies and especially their beat frequencies. The higher energetic part of experimentally measured incoherent losses has been identified as mainly prompt losses. The simulations using eigenmode structure given by \textsc{Ligka} revealed that the lowest energetic losses result from a combination of two interconnected facts: first, the many poloidal harmonics of the Toroidicity-induced Alfv{\'e}n eigenmode, which are caused by the gap alignment in the continuum of an inverted \qp\ equilibrium. Second, due to particle drift orbits that are radially extended on the size of the machine's small radius. In scenarios like this, especially in the presence of multiple modes with different frequencies, i.e.\ with resonances, that are complementary in phase space, a domino \cite{Berk95-III} effect can occur: particles in the high energy range leave the plasma as prompt losses, followed by resonant and diffusive losses in the lower energy region. At the same time, the redistribution caused by the core-localized mode refills the particles, that have been transported radially outwards by the outer mode. When the outer mode reaches the stochasticity threshold, low energy losses appear (down to 1/3 of the birth energy), and if many poloidal harmonics are present, they transport even very low energetic particles (down to 1/10 of the birth energy) across the loss boundary. This phase space channeling effect, caused by the presence of multiple modes, is clearly a nonlinear phenomenon, although the regime is still weakly nonlinear. Losses are enhanced by orders of magnitude. This result stresses the importance of the mode structure and thus infers a possible control of energetic particle transport via gap-dealignment by \qp\ and density shaping.\\
Another important result is the crucial role of the linearly subdominant mode for the nonlinear energetic particle transport. To follow up this finding, further studies are planned in the near future, investigating the nonlinear behaviour of a ``sea'' of linearly stable or weakly unstable modes. Such scenario is one of those considered most realistic for \textsc{Iter}.

\section*{Acknowledgment}
This work was facilitated by the Max-Planck/Princeton Center for Plasma Physics. 
The simulation results were obtained with the help of the high
performance computing resources provided by the RZG in Garching, Germany,
on the \textsc{Hpc-FF} system at Forschungszentrum J\"ulich, Germany, and on CSC's \textsc{Iferc-Helios} in Japan. Thanks. 

\bibliographystyle{apsrev}

\bibliography{literature}

\begin{thebibliography}{29}
\expandafter\ifx\csname natexlab\endcsname\relax\def\natexlab#1{#1}\fi
\expandafter\ifx\csname bibnamefont\endcsname\relax
  \def\bibnamefont#1{#1}\fi
\expandafter\ifx\csname bibfnamefont\endcsname\relax
  \def\bibfnamefont#1{#1}\fi
\expandafter\ifx\csname citenamefont\endcsname\relax
  \def\citenamefont#1{#1}\fi
\expandafter\ifx\csname url\endcsname\relax
  \def\url#1{\texttt{#1}}\fi
\expandafter\ifx\csname urlprefix\endcsname\relax\def\urlprefix{URL }\fi
\providecommand{\bibinfo}[2]{#2}
\providecommand{\eprint}[2][]{\url{#2}}

\bibitem[{\citenamefont{Pinches et~al.}(1998)\citenamefont{Pinches, Appel,
  Candy, Sharapov, Berk, Borba, Breizman, Hender, Hopcraft, Huysmans
  et~al.}}]{Pinches98}
\bibinfo{author}{\bibfnamefont{S.}~\bibnamefont{Pinches}},
  \bibinfo{author}{\bibfnamefont{L.}~\bibnamefont{Appel}},
  \bibinfo{author}{\bibfnamefont{J.}~\bibnamefont{Candy}},
  \bibinfo{author}{\bibfnamefont{S.}~\bibnamefont{Sharapov}},
  \bibinfo{author}{\bibfnamefont{H.}~\bibnamefont{Berk}},
  \bibinfo{author}{\bibfnamefont{D.}~\bibnamefont{Borba}},
  \bibinfo{author}{\bibfnamefont{B.}~\bibnamefont{Breizman}},
  \bibinfo{author}{\bibfnamefont{T.}~\bibnamefont{Hender}},
  \bibinfo{author}{\bibfnamefont{K.}~\bibnamefont{Hopcraft}},
  \bibinfo{author}{\bibfnamefont{G.}~\bibnamefont{Huysmans}},
  \bibnamefont{et~al.}, \emph{\bibinfo{title}{The {HAGIS} self-consistent
  nonlinear wave-particle interaction model}}, \bibinfo{journal}{Comput. Phys.
  Commun.} \textbf{\bibinfo{volume}{111}}, \bibinfo{pages}{133 }
  (\bibinfo{year}{1998}), ISSN \bibinfo{issn}{0010-4655},
  \urlprefix\url{http://www.sciencedirect.com/science/article/pii/S0010465598000344}.

\bibitem[{\citenamefont{Br{\"u}dgam}(2010)}]{mwb_phd}
\bibinfo{author}{\bibfnamefont{M.}~\bibnamefont{Br{\"u}dgam}}, Ph.D. thesis,
  \bibinfo{school}{Technische Universit{\"a}t M{\"u}nchen}
  (\bibinfo{year}{2010}).

\bibitem[{\citenamefont{{Lauber Ph} et~al.}(2007)\citenamefont{{Lauber Ph},
  G{\"u}nter, K{\"o}nies, and Pinches}}]{Lauber07}
\bibinfo{author}{\bibnamefont{{Lauber Ph}}},
  \bibinfo{author}{\bibfnamefont{S.}~\bibnamefont{G{\"u}nter}},
  \bibinfo{author}{\bibfnamefont{A.}~\bibnamefont{K{\"o}nies}},
  \bibnamefont{and} \bibinfo{author}{\bibfnamefont{S.~D.}
  \bibnamefont{Pinches}}, \emph{\bibinfo{title}{{LIGKA:} a linear gyrokinetic
  code for the description of background kinetic and fast particle effects on
  the {MHD} stability in tokamaks}}, \bibinfo{journal}{J. Comp. Phys.}
  \textbf{\bibinfo{volume}{226}}, \bibinfo{pages}{447 } (\bibinfo{year}{2007}),
  ISSN \bibinfo{issn}{0021-9991},
  \urlprefix\url{http://www.sciencedirect.com/science/article/pii/S0021999107001660}.

\bibitem[{\citenamefont{Garc{\'i}a-Mu{\~n}oz
  et~al.}(2010)\citenamefont{Garc{\'i}a-Mu{\~n}oz, Hicks, van Voornveld,
  Classen, Bilato, Bobkov, Br{\"u}dgam, Fahrbach, Igochine, Jaemsae
  et~al.}}]{Garcia10}
\bibinfo{author}{\bibfnamefont{M.}~\bibnamefont{Garc{\'i}a-Mu{\~n}oz}},
  \bibinfo{author}{\bibfnamefont{N.}~\bibnamefont{Hicks}},
  \bibinfo{author}{\bibfnamefont{R.}~\bibnamefont{van Voornveld}},
  \bibinfo{author}{\bibfnamefont{I.~G.~J.} \bibnamefont{Classen}},
  \bibinfo{author}{\bibfnamefont{R.}~\bibnamefont{Bilato}},
  \bibinfo{author}{\bibfnamefont{V.}~\bibnamefont{Bobkov}},
  \bibinfo{author}{\bibfnamefont{M.}~\bibnamefont{Br{\"u}dgam}},
  \bibinfo{author}{\bibfnamefont{H.-U.} \bibnamefont{Fahrbach}},
  \bibinfo{author}{\bibfnamefont{V.}~\bibnamefont{Igochine}},
  \bibinfo{author}{\bibfnamefont{S.}~\bibnamefont{Jaemsae}},
  \bibnamefont{et~al.}, \emph{\bibinfo{title}{Convective and diffusive
  energetic particle losses induced by {S}hear {Alfv{\'e}n} waves in the {ASDEX
  Upgrade} tokamak}}, \bibinfo{journal}{Phys. Rev. Lett.}
  \textbf{\bibinfo{volume}{104}}, \bibinfo{pages}{185002}
  (\bibinfo{year}{2010}),
  \urlprefix\url{http://link.aps.org/doi/10.1103/PhysRevLett.104.185002}.

\bibitem[{\citenamefont{Berk et~al.}(1995)\citenamefont{Berk, Breizman,
  Fitzpatrick, and Wong}}]{Berk95-III}
\bibinfo{author}{\bibfnamefont{H.}~\bibnamefont{Berk}},
  \bibinfo{author}{\bibfnamefont{B.}~\bibnamefont{Breizman}},
  \bibinfo{author}{\bibfnamefont{J.}~\bibnamefont{Fitzpatrick}},
  \bibnamefont{and} \bibinfo{author}{\bibfnamefont{H.}~\bibnamefont{Wong}},
  \emph{\bibinfo{title}{Line broadened quasi-linear burst model [fusion
  plasma]}}, \bibinfo{journal}{Nucl. Fusion} \textbf{\bibinfo{volume}{35}},
  \bibinfo{pages}{1661} (\bibinfo{year}{1995}),
  \urlprefix\url{http://stacks.iop.org/0029-5515/35/i=12/a=I30}.

\bibitem[{\citenamefont{Cheng et~al.}(1985)\citenamefont{Cheng, Chen, and
  Chance}}]{Cheng85}
\bibinfo{author}{\bibfnamefont{C.~Z.} \bibnamefont{Cheng}},
  \bibinfo{author}{\bibfnamefont{L.}~\bibnamefont{Chen}}, \bibnamefont{and}
  \bibinfo{author}{\bibfnamefont{M.}~\bibnamefont{Chance}},
  \emph{\bibinfo{title}{High-n ideal and resistive shear {Alfv{\'e}n} waves in
  tokamaks}}, \bibinfo{journal}{Ann. Phys.} \textbf{\bibinfo{volume}{161}},
  \bibinfo{pages}{21 } (\bibinfo{year}{1985}),
  \urlprefix\url{http://www.sciencedirect.com/science/article/pii/0003491685903355}.

\bibitem[{\citenamefont{Cheng and Chance}(1986)}]{Cheng86}
\bibinfo{author}{\bibfnamefont{C.~Z.} \bibnamefont{Cheng}} \bibnamefont{and}
  \bibinfo{author}{\bibfnamefont{M.~S.} \bibnamefont{Chance}},
  \emph{\bibinfo{title}{Low-n shear {Alfv{\'e}n} spectra in axisymmetric
  toroidal plasmas}}, \bibinfo{journal}{Phys. Fluids}
  \textbf{\bibinfo{volume}{29}}, \bibinfo{pages}{3695} (\bibinfo{year}{1986}),
  \urlprefix\url{http://link.aip.org/link/?PFL/29/3695/1}.

\bibitem[{\citenamefont{Berk et~al.}(2001)\citenamefont{Berk, Borba, Breizman,
  Pinches, and Sharapov}}]{Berk01}
\bibinfo{author}{\bibfnamefont{H.~L.} \bibnamefont{Berk}},
  \bibinfo{author}{\bibfnamefont{D.~N.} \bibnamefont{Borba}},
  \bibinfo{author}{\bibfnamefont{B.~N.} \bibnamefont{Breizman}},
  \bibinfo{author}{\bibfnamefont{S.~D.} \bibnamefont{Pinches}},
  \bibnamefont{and} \bibinfo{author}{\bibfnamefont{S.~E.}
  \bibnamefont{Sharapov}}, \emph{\bibinfo{title}{Theoretical interpretation of
  {Alfv{\'e}n} cascades in tokamaks with nonmonotonic $q$ profiles}},
  \bibinfo{journal}{Phys. Rev. Lett.} \textbf{\bibinfo{volume}{87}}
  (\bibinfo{year}{2001}),
  \urlprefix\url{http://link.aps.org/doi/10.1103/PhysRevLett.87.185002}.

\bibitem[{\citenamefont{Breizman et~al.}(2003)\citenamefont{Breizman, Berk,
  Pekker, Pinches, and Sharapov}}]{Breizman03}
\bibinfo{author}{\bibfnamefont{B.~N.} \bibnamefont{Breizman}},
  \bibinfo{author}{\bibfnamefont{H.~L.} \bibnamefont{Berk}},
  \bibinfo{author}{\bibfnamefont{M.~S.} \bibnamefont{Pekker}},
  \bibinfo{author}{\bibfnamefont{S.~D.} \bibnamefont{Pinches}},
  \bibnamefont{and} \bibinfo{author}{\bibfnamefont{S.~E.}
  \bibnamefont{Sharapov}}, \emph{\bibinfo{title}{Theory of {A}lfv{\'e}n
  eigenmodes in shear reversed plasmas}}, \bibinfo{journal}{Phys. Plasmas}
  \textbf{\bibinfo{volume}{10}}, \bibinfo{pages}{3649} (\bibinfo{year}{2003}),
  \urlprefix\url{http://link.aip.org/link/?PHP/10/3649/1}.

\bibitem[{\citenamefont{Heidbrink et~al.}(1993)\citenamefont{Heidbrink, Strait,
  Chu, and Turnbull}}]{Heidbrink93}
\bibinfo{author}{\bibfnamefont{W.~W.} \bibnamefont{Heidbrink}},
  \bibinfo{author}{\bibfnamefont{E.~J.} \bibnamefont{Strait}},
  \bibinfo{author}{\bibfnamefont{M.~S.} \bibnamefont{Chu}}, \bibnamefont{and}
  \bibinfo{author}{\bibfnamefont{A.~D.} \bibnamefont{Turnbull}},
  \emph{\bibinfo{title}{Observation of beta-induced {A}lfv{\'e}n eigenmodes in
  the {DIII-D} tokamak}}, \bibinfo{journal}{Phys. Rev. Lett.}
  \textbf{\bibinfo{volume}{71}}, \bibinfo{pages}{855} (\bibinfo{year}{1993}),
  \urlprefix\url{http://link.aps.org/doi/10.1103/PhysRevLett.71.855}.

\bibitem[{\citenamefont{Turnbull et~al.}(1993)\citenamefont{Turnbull, Strait,
  Heidbrink, Chu, Duong, Greene, Lao, Taylor, and Thompson}}]{Turnbull93}
\bibinfo{author}{\bibfnamefont{A.~D.} \bibnamefont{Turnbull}},
  \bibinfo{author}{\bibfnamefont{E.~J.} \bibnamefont{Strait}},
  \bibinfo{author}{\bibfnamefont{W.~W.} \bibnamefont{Heidbrink}},
  \bibinfo{author}{\bibfnamefont{M.~S.} \bibnamefont{Chu}},
  \bibinfo{author}{\bibfnamefont{H.~H.} \bibnamefont{Duong}},
  \bibinfo{author}{\bibfnamefont{J.~M.} \bibnamefont{Greene}},
  \bibinfo{author}{\bibfnamefont{L.~L.} \bibnamefont{Lao}},
  \bibinfo{author}{\bibfnamefont{T.~S.} \bibnamefont{Taylor}},
  \bibnamefont{and} \bibinfo{author}{\bibfnamefont{S.~J.}
  \bibnamefont{Thompson}}, \emph{\bibinfo{title}{Global {A}lfv{\'e}n modes:
  Theory and experiment}}, \bibinfo{journal}{Phys. Fluids B}
  \textbf{\bibinfo{volume}{5}}, \bibinfo{pages}{2546} (\bibinfo{year}{1993}),
  \urlprefix\url{http://link.aip.org/link/?PFB/5/2546/1}.

\bibitem[{\citenamefont{Gruber et~al.}(1984)\citenamefont{Gruber, Kaufmann,
  K{\"o}ppend{\"o}rfer, Lackner, and Neuhauser}}]{Gruber84}
\bibinfo{author}{\bibfnamefont{O.}~\bibnamefont{Gruber}},
  \bibinfo{author}{\bibfnamefont{M.}~\bibnamefont{Kaufmann}},
  \bibinfo{author}{\bibfnamefont{W.}~\bibnamefont{K{\"o}ppend{\"o}rfer}},
  \bibinfo{author}{\bibfnamefont{K.}~\bibnamefont{Lackner}}, \bibnamefont{and}
  \bibinfo{author}{\bibfnamefont{J.}~\bibnamefont{Neuhauser}},
  \emph{\bibinfo{title}{Physics background of the {ASDEX Upgrade} project}},
  \bibinfo{journal}{Journal of Nuclear Materials}
  \textbf{\bibinfo{volume}{121}}, \bibinfo{pages}{407} (\bibinfo{year}{1984}),
  \urlprefix\url{http://www.sciencedirect.com/science/article/pii/0022311584901533}.

\bibitem[{\citenamefont{Garc{\'i}a-Mu{\~n}oz
  et~al.}(2009{\natexlab{a}})\citenamefont{Garc{\'i}a-Mu{\~n}oz, Fahrbach,
  Zohm, and {the ASDEX Upgrade Team}}}]{GarciaRev09}
\bibinfo{author}{\bibfnamefont{M.}~\bibnamefont{Garc{\'i}a-Mu{\~n}oz}},
  \bibinfo{author}{\bibfnamefont{H.-U.} \bibnamefont{Fahrbach}},
  \bibinfo{author}{\bibfnamefont{H.}~\bibnamefont{Zohm}}, \bibnamefont{and}
  \bibinfo{author}{\bibnamefont{{the ASDEX Upgrade Team}}},
  \emph{\bibinfo{title}{Scintillator based detector for fast-ion losses induced
  by magnetohydrodynamic instabilities in the {ASDEX Upgrade} tokamak}},
  \bibinfo{journal}{Rev. Sci. Instrum.} \textbf{\bibinfo{volume}{80}},
  \bibinfo{eid}{053503} (\bibinfo{year}{2009}{\natexlab{a}}),
  \urlprefix\url{http://link.aip.org/link/?RSI/80/053503/1}.

\bibitem[{\citenamefont{Zeeland et~al.}(2011)\citenamefont{Zeeland, Heidbrink,
  Fisher, Garc{\'i}a-Mu{\~n}oz, Kramer, Pace, White, Aekaeslompolo, Austin,
  Boom et~al.}}]{zeeland11}
\bibinfo{author}{\bibfnamefont{M.~A.~V.} \bibnamefont{Zeeland}},
  \bibinfo{author}{\bibfnamefont{W.~W.} \bibnamefont{Heidbrink}},
  \bibinfo{author}{\bibfnamefont{R.~K.} \bibnamefont{Fisher}},
  \bibinfo{author}{\bibfnamefont{M.~M.} \bibnamefont{Garc{\'i}a-Mu{\~n}oz}},
  \bibinfo{author}{\bibfnamefont{G.~J.} \bibnamefont{Kramer}},
  \bibinfo{author}{\bibfnamefont{D.~C.} \bibnamefont{Pace}},
  \bibinfo{author}{\bibfnamefont{R.~B.} \bibnamefont{White}},
  \bibinfo{author}{\bibfnamefont{S.}~\bibnamefont{Aekaeslompolo}},
  \bibinfo{author}{\bibfnamefont{M.~E.} \bibnamefont{Austin}},
  \bibinfo{author}{\bibfnamefont{J.~E.} \bibnamefont{Boom}},
  \bibnamefont{et~al.}, \emph{\bibinfo{title}{Measurements and modeling of
  {Alfv{\'e}n} eigenmode induced fast ion transport and loss in {DIII-D} and
  {ASDEX Upgrade}}}, \bibinfo{journal}{Phys. Plasmas}
  \textbf{\bibinfo{volume}{18}}, \bibinfo{pages}{056114}
  (\bibinfo{year}{2011}),
  \urlprefix\url{http://link.aip.org/link/?PHP/18/056114/1}.

\bibitem[{\citenamefont{Sigmar et~al.}(1992)\citenamefont{Sigmar, Hsu, White,
  and Cheng}}]{Sigmar92}
\bibinfo{author}{\bibfnamefont{D.}~\bibnamefont{Sigmar}},
  \bibinfo{author}{\bibfnamefont{C.}~\bibnamefont{Hsu}},
  \bibinfo{author}{\bibfnamefont{R.}~\bibnamefont{White}}, \bibnamefont{and}
  \bibinfo{author}{\bibfnamefont{C.}~\bibnamefont{Cheng}},
  \emph{\bibinfo{title}{Alpha-particle losses from toroidicity-induced
  {A}lfv{\'e}n eigenmodes. {Part II}: {Monte Carlo} simulations and anomalous
  alpha-loss processes}}, \bibinfo{journal}{Phys. Fluids B}
  \textbf{\bibinfo{volume}{4}}, \bibinfo{pages}{1506} (\bibinfo{year}{1992}).

\bibitem[{\citenamefont{Pinches}(1996)}]{sip_phd}
\bibinfo{author}{\bibfnamefont{S.~D.} \bibnamefont{Pinches}}, Ph.D. thesis,
  \bibinfo{school}{University of Nottingham} (\bibinfo{year}{1996}),
  \urlprefix\url{http://www.ipp.mpg.de/~Simon.Pinches/thesis/thesis.html}.

\bibitem[{\citenamefont{{Huysmans G. T. A. et al}}(1990)}]{Huysmans91}
\bibinfo{author}{\bibnamefont{{Huysmans G. T. A. et al}}},
  \bibinfo{organization}{Conf. on Computational Physics Proc. World Scientific,
  Singapore} (\bibinfo{publisher}{AIP}, \bibinfo{year}{1990}).

\bibitem[{\citenamefont{Igochine et~al.}(2003)\citenamefont{Igochine,
  G{\"u}nter, Maraschek, and {the ASDEX Upgrade Team}}}]{Igochine03}
\bibinfo{author}{\bibfnamefont{V.}~\bibnamefont{Igochine}},
  \bibinfo{author}{\bibfnamefont{S.}~\bibnamefont{G{\"u}nter}},
  \bibinfo{author}{\bibfnamefont{M.}~\bibnamefont{Maraschek}},
  \bibnamefont{and} \bibinfo{author}{\bibnamefont{{the ASDEX Upgrade Team}}},
  \emph{\bibinfo{title}{Investigation of complex {MHD} activity by a combined
  use of various diagnostics}}, \bibinfo{journal}{Nucl. Fusion}
  \textbf{\bibinfo{volume}{43}}, \bibinfo{pages}{1801} (\bibinfo{year}{2003}),
  \urlprefix\url{http://stacks.iop.org/0029-5515/43/i=12/a=023}.

\bibitem[{\citenamefont{{Mc Carthy P. J.} and {the ASDEX Upgrade
  Team}}(2012)}]{Carthy12}
\bibinfo{author}{\bibnamefont{{Mc Carthy P. J.}}} \bibnamefont{and}
  \bibinfo{author}{\bibnamefont{{the ASDEX Upgrade Team}}},
  \emph{\bibinfo{title}{Identification of edge-localized moments of the current
  density profile in a tokamak equilibrium from external magnetic
  measurements}}, \bibinfo{journal}{Plasma Phys. Control. Fusion}
  \textbf{\bibinfo{volume}{54}}, \bibinfo{pages}{015010}
  (\bibinfo{year}{2012}),
  \urlprefix\url{http://stacks.iop.org/0741-3335/54/i=1/a=015010}.

\bibitem[{\citenamefont{{Lauber Ph} et~al.}(2009)\citenamefont{{Lauber Ph},
  Br{\"u}dgam, Curran, Igochine, Sassenberg, G{\"u}nter, Maraschek,
  Garc{\'i}a-Mu{\~n}oz, Hicks, and {the ASDEX Upgrade Team}}}]{Lauber09}
\bibinfo{author}{\bibnamefont{{Lauber Ph}}},
  \bibinfo{author}{\bibfnamefont{M.}~\bibnamefont{Br{\"u}dgam}},
  \bibinfo{author}{\bibfnamefont{D.}~\bibnamefont{Curran}},
  \bibinfo{author}{\bibfnamefont{V.}~\bibnamefont{Igochine}},
  \bibinfo{author}{\bibfnamefont{K.}~\bibnamefont{Sassenberg}},
  \bibinfo{author}{\bibfnamefont{S.}~\bibnamefont{G{\"u}nter}},
  \bibinfo{author}{\bibfnamefont{M.}~\bibnamefont{Maraschek}},
  \bibinfo{author}{\bibfnamefont{M.}~\bibnamefont{Garc{\'i}a-Mu{\~n}oz}},
  \bibinfo{author}{\bibfnamefont{N.}~\bibnamefont{Hicks}}, \bibnamefont{and}
  \bibinfo{author}{\bibnamefont{{the ASDEX Upgrade Team}}},
  \emph{\bibinfo{title}{Kinetic {{A}lfv{\'e}n} eigenmodes at {ASDEX Upgrade}}},
  \bibinfo{journal}{Plasma Phys. Control. Fusion}
  \textbf{\bibinfo{volume}{51}}, \bibinfo{pages}{124009}
  (\bibinfo{year}{2009}),
  \urlprefix\url{http://stacks.iop.org/0741-3335/51/i=12/a=124009}.

\bibitem[{\citenamefont{Porcelli et~al.}(1994)\citenamefont{Porcelli,
  Stankiewicz, Kerner, and Berk}}]{porcelli94}
\bibinfo{author}{\bibfnamefont{F.}~\bibnamefont{Porcelli}},
  \bibinfo{author}{\bibfnamefont{R.}~\bibnamefont{Stankiewicz}},
  \bibinfo{author}{\bibfnamefont{W.}~\bibnamefont{Kerner}}, \bibnamefont{and}
  \bibinfo{author}{\bibfnamefont{H.~L.} \bibnamefont{Berk}},
  \emph{\bibinfo{title}{Solution of the drift-kinetic equation for global
  plasma modes and finite particle orbit widths}}, \bibinfo{journal}{Phys.
  Plasmas} \textbf{\bibinfo{volume}{1}}, \bibinfo{pages}{470}
  (\bibinfo{year}{1994}),
  \urlprefix\url{http://link.aip.org/link/?PHP/1/470/1}.

\bibitem[{\citenamefont{Pinches et~al.}(2006)\citenamefont{Pinches, Kiptily,
  Sharapov, Darrow, Eriksson, Fahrbach, Garc{{\'i}a}-Mu{\~n}oz, Reich,
  Strumberger, Werner et~al.}}]{Pinches06}
\bibinfo{author}{\bibfnamefont{S.~D.} \bibnamefont{Pinches}},
  \bibinfo{author}{\bibfnamefont{V.}~\bibnamefont{Kiptily}},
  \bibinfo{author}{\bibfnamefont{S.}~\bibnamefont{Sharapov}},
  \bibinfo{author}{\bibfnamefont{D.}~\bibnamefont{Darrow}},
  \bibinfo{author}{\bibfnamefont{L.-G.} \bibnamefont{Eriksson}},
  \bibinfo{author}{\bibfnamefont{H.-U.} \bibnamefont{Fahrbach}},
  \bibinfo{author}{\bibfnamefont{M.}~\bibnamefont{Garc{{\'i}a}-Mu{\~n}oz}},
  \bibinfo{author}{\bibfnamefont{M.}~\bibnamefont{Reich}},
  \bibinfo{author}{\bibfnamefont{E.}~\bibnamefont{Strumberger}},
  \bibinfo{author}{\bibfnamefont{A.}~\bibnamefont{Werner}},
  \bibnamefont{et~al.}, \emph{\bibinfo{title}{Observation and modelling of fast
  ion loss in {JET} and {ASDEX Upgrade}}}, \bibinfo{journal}{Nucl. Fusion}
  \textbf{\bibinfo{volume}{46}}, \bibinfo{pages}{S904} (\bibinfo{year}{2006}),
  \urlprefix\url{http://stacks.iop.org/0029-5515/46/i=10/a=S06}.

\bibitem[{\citenamefont{Schneller}(2013)}]{mirjam_phd}
\bibinfo{author}{\bibfnamefont{M.}~\bibnamefont{Schneller}}, Ph.D. thesis,
  \bibinfo{school}{Technische Universit{\"a}t M{\"u}nchen}
  (\bibinfo{year}{2013}).

\bibitem[{\citenamefont{Bilato et~al.}(2012)\citenamefont{Bilato, Brambilla,
  and Jiang}}]{Bilato12}
\bibinfo{author}{\bibfnamefont{R.}~\bibnamefont{Bilato}},
  \bibinfo{author}{\bibfnamefont{M.}~\bibnamefont{Brambilla}},
  \bibnamefont{and} \bibinfo{author}{\bibfnamefont{Z.}~\bibnamefont{Jiang}},
  \emph{\bibinfo{title}{Implementing zero-banana-width quasilinear operator for
  fast {ICRF} simulations}}, \bibinfo{journal}{Journal of Physics: Conference
  Series} \textbf{\bibinfo{volume}{401}}, \bibinfo{pages}{012001}
  (\bibinfo{year}{2012}),
  \urlprefix\url{http://stacks.iop.org/1742-6596/401/i=1/a=012001}.

\bibitem[{\citenamefont{Zonca and Chen}(2000)}]{Zonca00}
\bibinfo{author}{\bibfnamefont{F.}~\bibnamefont{Zonca}} \bibnamefont{and}
  \bibinfo{author}{\bibfnamefont{L.}~\bibnamefont{Chen}},
  \emph{\bibinfo{title}{Destabilization of energetic particle modes by
  localized particle sources}}, \bibinfo{journal}{Phys. Plasmas}
  \textbf{\bibinfo{volume}{7}}, \bibinfo{pages}{4600} (\bibinfo{year}{2000}),
  \urlprefix\url{http://link.aip.org/link/?PHP/7/4600/1}.

\bibitem[{\citenamefont{Troia}(2012)}]{Troia12}
\bibinfo{author}{\bibfnamefont{C.~D.} \bibnamefont{Troia}},
  \emph{\bibinfo{title}{From the orbit theory to a guiding center parametric
  equilibrium distribution function}}, \bibinfo{journal}{Plasma Phys. Control.
  Fusion} \textbf{\bibinfo{volume}{54}}, \bibinfo{pages}{105017}
  (\bibinfo{year}{2012}),
  \urlprefix\url{http://stacks.iop.org/0741-3335/54/i=10/a=105017}.

\bibitem[{\citenamefont{{Curran D, Schneider W}}()}]{Diarmuid+Schneider-priv}
\bibinfo{author}{\bibnamefont{{Curran D, Schneider W}}},
  \emph{\bibinfo{title}{private communication}}.

\bibitem[{\citenamefont{Schneller et~al.}(2012)\citenamefont{Schneller, {Lauber
  Ph}, Br{\"u}dgam, Pinches, and G{\"u}nter}}]{Schneller12}
\bibinfo{author}{\bibfnamefont{M.}~\bibnamefont{Schneller}},
  \bibinfo{author}{\bibnamefont{{Lauber Ph}}},
  \bibinfo{author}{\bibfnamefont{M.}~\bibnamefont{Br{\"u}dgam}},
  \bibinfo{author}{\bibfnamefont{S.~D.} \bibnamefont{Pinches}},
  \bibnamefont{and}
  \bibinfo{author}{\bibfnamefont{S.}~\bibnamefont{G{\"u}nter}},
  \emph{\bibinfo{title}{Double-resonant fast particle-wave interaction}},
  \bibinfo{journal}{Nuclear Fusion} \textbf{\bibinfo{volume}{52}},
  \bibinfo{pages}{103019} (\bibinfo{year}{2012}),
  \urlprefix\url{http://stacks.iop.org/0029-5515/52/i=10/a=103019}.

\bibitem[{\citenamefont{Garc{\'i}a-Mu{\~n}oz
  et~al.}(2009{\natexlab{b}})\citenamefont{Garc{\'i}a-Mu{\~n}oz, Fahrbach,
  Bobkov, Hicks, Igochine, Jaemsae, Maraschek, Sassenberg, and {the ASDEX
  Upgrade Team}}}]{GarciaIAEA09}
\bibinfo{author}{\bibfnamefont{M.}~\bibnamefont{Garc{\'i}a-Mu{\~n}oz}},
  \bibinfo{author}{\bibfnamefont{H.-U.} \bibnamefont{Fahrbach}},
  \bibinfo{author}{\bibfnamefont{V.}~\bibnamefont{Bobkov}},
  \bibinfo{author}{\bibfnamefont{N.}~\bibnamefont{Hicks}},
  \bibinfo{author}{\bibfnamefont{V.}~\bibnamefont{Igochine}},
  \bibinfo{author}{\bibfnamefont{S.}~\bibnamefont{Jaemsae}},
  \bibinfo{author}{\bibfnamefont{M.}~\bibnamefont{Maraschek}},
  \bibinfo{author}{\bibfnamefont{K.}~\bibnamefont{Sassenberg}},
  \bibnamefont{and} \bibinfo{author}{\bibnamefont{{the ASDEX Upgrade Team}}},
  \bibinfo{organization}{11th {IAEA} Technical Meeting on Energetic Particles
  in Magnetic Confinement Systems} (\bibinfo{publisher}{International Atomic
  Energy Agency}, \bibinfo{address}{Kiev}, \bibinfo{year}{2009}{\natexlab{b}}).

\end{thebibliography}

\end{document}